\documentclass[aip,pop,reprint,amsmath,amssymb,floatfix]{revtex4-1}

\pdfoutput=1

\usepackage{graphicx}
\usepackage{subfigure}

\let\oldhat\hat
\renewcommand{\vec}[1]{\mathbf{#1}}
\renewcommand{\hat}[1]{\oldhat{\mathbf{#1}}}

\begin{document}

\title{On the spatio-temporal behavior of
magnetohydrodynamic turbulence in a magnetized plasma}

\author{R. Lugones}\email{rlugones@df.uba.ar}
\affiliation{Departamento de F\'\i sica, Facultad de Ciencias Exactas y Naturales,
  Universidad de Buenos Aires and IFIBA, CONICET, Ciudad
  Universitaria, 1428 Buenos Aires, Argentina.}

\author{P. Dmitruk}
\affiliation{Departamento de F\'\i sica, Facultad de Ciencias Exactas y Naturales,
  Universidad de Buenos Aires and IFIBA, CONICET, Ciudad
  Universitaria, 1428 Buenos Aires, Argentina.}

\author{P.D. Mininni}
\affiliation{Departamento de F\'\i sica, Facultad de Ciencias Exactas y Naturales,
  Universidad de Buenos Aires and IFIBA, CONICET, Ciudad
  Universitaria, 1428 Buenos Aires, Argentina.}

\author{M. Wan}
\affiliation{Bartol Research Institute and Department of Physics and Astronomy,
  University of Delaware, Newark, DE 19716, USA.}

\author{W.H. Matthaeus}
\affiliation{Bartol Research Institute and Department of Physics and Astronomy,
  University of Delaware, Newark, DE 19716, USA.}

\begin{abstract}
  Using direct numerical simulations of three-dimensional magnetohydrodynamic (MHD)
  turbulence the spatio-temporal behavior of magnetic field
  fluctuations is analyzed.  Cases with relatively small, medium and
  large values of a mean background magnetic field are considered. The
  (wavenumber) scale dependent time correlation function is directly
  computed for different simulations, varying the mean magnetic field
  value. From this correlation function the time decorrelation is
  computed and compared with different theoretical times, namely,
  the local non-linear time, the random sweeping time, and the
  Alfv\'enic time, the latter being a wave effect. It is observed that
  time decorrelations are dominated by sweeping effects, and only at
  large values of the mean magnetic field and for wave vectors mainly
  aligned with this field time decorrelations are controlled by
  Alfv\'enic effects.
\end{abstract}

\maketitle

\section{Introduction}\label{sec_Intro}
 
It is known that in the linear approximation the magnetohydrodynamic
(MHD) equations can sustain Alfv\'en waves. The simplest case
corresponds to incompressible MHD with a uniform background magnetic
field ${\bf  B_0}$, for which the linear dispersion relation (in the ideal
non-dissipative case) describes waves with frequency $w={\bf k} \cdot
{\bf v_A}$, for wavevector ${\bf k}$, Alfv\'en velocity ${\bf
  v_A}={\bf B_0}/\sqrt{4\pi \rho}$, and density $\rho$. Also, the
complex Fourier components of the velocity ${\bf v}({\bf k})$ and
of magnetic field fluctuations ${\bf b}({\bf k})$ are transverse to
the wavevector, i.e., ${\bf v}({\bf k}) \cdot {\bf k} = {\bf b} \cdot
{\bf k}=0$. Interestingly, these waves when considered in isolation
are exact nonlinear solutions of the MHD equations.

However, when non-linear terms are taken into account, the system can
also develop far from equilibrium dynamics, with the waves coexisting
with eddies in a fully developed turbulent flow
\cite{dmitruk_waves_2009}. In this turbulent regime one does not
necessarily expect a direct or explicit relation between frequency and
wavenumber, such as the dispersion relation for waves. This regime is
characterized by interactions of several types, such as local-in-scale
nonlinear distortion of eddies \cite{monin_statistical_2013,
  kolmogorov_local_1941, mccomb_physics_1992}, and non-local effects
\cite{alexakis_turbulent_2007, alexakis_anisotropic_2007,
  teaca_energy_2009, mininni_scale_2011} the most extreme of which is
transport or ``sweeping'' of small eddies by large eddies
\cite{kraichnan_structure_1959, tennekes_eulerian_1975,
  chen_sweeping_1989, nelkin_time_1990}. Furthermore, for MHD
turbulence \cite{pouquet_strong_1976, zhou_magnetohydrodynamic_2004},
in addition to the global nonlinear time $\tau_{nl}$, there are also
time scales associated with scale-dependent (local) nonlinear effects,
nonlocal sweeping, and wave propagation
\cite{zhou_magnetohydrodynamic_2004}.

In the early 70's, investigation of hydrodynamic turbulence was
directed to study the decorrelation time of the velocity field
\cite{orszag_numerical_1972, orszag_analytical_1970,
  tennekes_eulerian_1975, heisenberg_zur_1948,
  comte-bellot_simple_1971}.  The main conclusion was that the
sweeping dominates the temporal decorrelation in the inertial range
\cite{zhou_non-gaussian_1993, sanada_random_1992}. Recently, a similar
study has been implemented in magnetohydrodynamics
\cite{servidio_time_2011, matthaeus_eulerian_2010,
  carbone_anisotropy_2011}. One difference with the hydrodynamic case
is the presence of other non-local phenomena (besides the sweeping),
such as the Alfv\'enic propagation or Alfv\'enic distortion, namely
``magnetic sweeping''. The main result of Servidio {\it et al.}
\cite{servidio_time_2011} on the temporal decorrelation for isotropic
turbulence was that, as in hydrodynamics, the temporal decorrelation
in MHD is governed by nonlocal interactions (in this case, sweeping
and Alfv\'en decorrelation). However, they were not able to
distinguish between the effect of sweeping and Alfv\'enic distortion.
In this paper, our main objective is to extend this analysis and
generalize it to magnetized plasmas at large scales where the MHD
approximation is valid.

In this work we study the different decorrelation times through the
various scales in the inertial range for MHD turbulence with a guide
field. The main objective is to understand the temporal decorrelation
of the fluctuations, by studying the relative value of decorrelation
times for the different scales. Thus, we will be able to relate the
scaling laws of the decorrelation times with the different
contributing physical effects: non-linear distorsion, random sweeping
and Alfv\'en wave propagation. In other words, we will study the
characteristic memory timescale for each spatial scale, in order to
identify the mechanisms of temporal decorrelation and to see whether
they are local or non-local. For this purpose, we will consider the
fluctuations at more than one length scale, to discern between the
different phenomena that are associated with temporal decorrelation,
in particular Alfv\'en wave propagation and random sweeping. This
method, based on the computation of spatio-temporal spectra and on
correlation functions, was proposed and implemented in rotating fluids
by Clark di Leoni {\it et al.}
\cite{clark_di_leoni_quantification_2014} (see also
\cite{clark_di_leoni_spatio-temporal_2015} for a general description
of the method). Meyrand and Galtier recently used the spatio-temporal
spectrum to study the transition from weak to strong turbulence in MHD
\cite{meyrand_direct_2016}, and intermittency in weak MHD turbulence
\cite{meyrand_weak_2015}. Here we consider the strong turbulent
regime, and compute both spectra as well as decorrelation times.

\section{Equations and numerical simulations}\label{sec_EqNumSim}

\subsection{The MHD equations}\label{sec_eq}

The incompressible MHD equations (momentum and induction
equations) in dimensionless units are
\begin{equation}\label{eq:MHD_v}
  \frac {\partial {\bf v}}{\partial t} +
  {\bf v }\cdot \nabla {\bf v} = -\frac{1}{\rho}\nabla p +
  {\bf j} \times {\bf B} + \frac{1}{R} \nabla^2{\bf v},
\end{equation}
\begin{equation}\label{eq:MHD_b}
  \frac{\partial {\bf b}}{\partial t} = \nabla \times ({\bf v} \times {\bf B})
  + \frac{1}{R_m} \nabla^2 {\bf b},
\end{equation}
where ${\bf v}$ is the plasma velocity; ${\bf B} = {\bf b} + {\bf
  B_0}$ the magnetic field, with a fluctuating part ${\bf b}$ and a
mean DC field ${\bf B_0}=B_0\hat{x}$; ${\bf j} = \nabla \times {\bf b}$ the
current density; $p$ the pressure, and $\rho$ the plasma density.  The
units are based on a characteristic speed $v_0$, which for MHD is
chosen to be the typical Alfv\'en speed of the magnetic field
fluctuations, $v_0 = \sqrt{\langle b^2 \rangle /(4\pi\rho)}$,
where $\langle . \rangle$ denotes a spatial average. The dimensionless
parameters appearing in the equations are the kinetic and magnetic
Reynolds numbers, $R=v_0 L/\nu$ and $R_m = v_0 L /\mu$ respectively, with $\nu$ the
kinematic viscosity, $\mu$ the magnetic diffusivity and $L$ the
characteristic length scale (the simulation box side length is defined
as $2\pi L$). The unit time is $t_0 = L/v_0$, which for MHD becomes
the Alfv\'en crossing time based on magnetic field fluctuations.

\subsection{Wavenumber-frequency spectrum and correlation functions}\label{sec_Wfspectrum_and_Gamma}

From Eqs. (\ref{eq:MHD_v}-\ref{eq:MHD_b}) and simple scaling arguments,
one can estimate different characteristic times. The local eddy
turnover time can be defined as $\tau_{nl} \sim \left[ k v(k)
\right]^{-1}$, where $k$ is the wave number and $v(k)$ is the amplitude
of velocity due to fluctuations at scale $\sim 1/k$. For a Kolmogorov-type
prediction of the velocity scaling, $v \sim v_{rms} \left(kL\right)^{-1/3}$,
the nonlinear time scales in the inertial range can be approximately
written as $\tau_{nl} = C_{nl} \left [ v_{rms} L^{-1/3}
  \left(\sqrt{k^2_\perp + k^2_\parallel}\right)^{2/3}\right ]^{-1}$,
where $C_{nl}$ is a dimensionless constant of order unity. In the
latter, $v_{rms} = \left\langle |{\bf v}|^2 \right\rangle ^{1/2}$ is a
global quantity, typically dominated by contributions from the large
scales.

The physics of time decorrelation depends on other effects and
therefore other available MHD time scales. One example is the sweeping
characteristic time at scale $\sim 1/k$, which may be expressed as
$\tau_{sw} = C_{sw} \left( v_{rms}\sqrt{k^2_\perp + k^2_\parallel}
\right)^{-1}$. This time corresponds to the advection of small scale
structures by the large scale flow. Analogously, a characteristic
Alfv\'en time can be defined as $\tau_A= C_A \left( B_0 k_\parallel
\right)^{-1}$. Here, $C_{sw}$ and $C_A$ are other dimensionless
constants of order unity. All these timescales depend on the wave
vector, and assuming the shortest timescale dominates the dynamics,
different regions in $k$-space in the energy spectrum can be defined.

The statistics of, for example, the magnetic field may be
characterized by the spatio-temporal two-point autocorrelation
function 
\begin{equation}
R({\bf r},\tau) = \left\langle {\bf b}( {\bf x},t) \cdot
  {\bf b}( {\bf x} + {\bf r},t+\tau) \right\rangle / \left\langle {\bf
    b}^2 \right\rangle.
\label{eq:Rbij}
\end{equation}
Note that this expression contains both the energy spectrum and the
Eulerian frequency spectrum (Wiener-Khinchin theorem); however, it
contains much more information which allows us to make a more subtle
analysis of the spatio-temporal relations. Fourier transforming in $r$
leads to a time-lagged spectral density which may be further
factorized as $S({\bf k},\tau) = S({\bf k})\Gamma({\bf k},t)$, where
${\bf k}$ is the wave vector. The function $\Gamma({\bf k},\tau)$, the
scale-dependent (or filtered) correlation function
\cite{heisenberg_zur_1948, comte-bellot_simple_1971,
  orszag_numerical_1972}, represents the dynamical decorrelation
effects describing the time decorrelation of each spatial mode 
${\bf k}$.

The function $\Gamma({\bf k},\tau)$ is thus the temporal correlation
function of the Fourier mode ${\bf k}$. Using this, we will be able to
identify the characteristic decorrelation time for each mode ${\bf k}$ and
therefore the loss of memory of 3D-fluctuations whose characteristics
lengths are of order $k_x^{-1}$, $k_y^{-1}$ and $k_z^{-1}$. When there
is no guide field we usually expect the flow to be isotropic both in real
space and in Fourier space, and therefore it is sufficient to study the
function $\Gamma(k,\tau)$ that depends only on $k=|{\bf k}|$. On the
other hand, in the presence of a guide field, the turbulence is
anisotropic; therefore, it is reasonable to use $\Gamma =
\Gamma(k_\perp,k_\parallel,\tau)$ where $k_\perp$ and $k_\parallel$
are the perpendicular and parallel (to the mean magnetic field) Fourier
wave numbers.

The function $\Gamma(k_\perp,k_\parallel,\tau)$ can help us to
understand the dynamics of different regions in Fourier space. For
example, the function $\Gamma(k_\perp=0,k_\parallel,\tau)$ give us
information about fluctuations that vary only in the parallel
direction. In the same way $\Gamma(k_\perp,k_\parallel=0,\tau)$ gives
information about fluctuations that vary only 
in the perpendicular direction. Also of interest is 
the information obtained from the
$\Gamma(k_\perp=k_0,k_\parallel,\tau)$ and the 
$\Gamma(k_\perp,k_\parallel=k_0,\tau)$ functions, when one of the
Fourier wavenumbers (the parallel or the perpendicular) is set to a
fixed value $k_0$. For example, studying the decorrelation time for
$\Gamma(k_\perp=k_0,k_\parallel,\tau)$ as a function of $k_\parallel$
would be useful to see the memory loss over time of 
fluctuations whose perpendicular characteristic length is 
$\sim k_0^{-1}$ (a fixed selected length), as a function of its
parallel scale $\sim k_\parallel^{-1}$. This would give us information
on two important issues: how the memory in one direction affects the
other, and more importantly, how to distinguish between random
sweeping and Alfv\'en propagation.

\subsection{Numerical simulations}\label{sec_NumSim}

We use a standard pseudospectral code to solve numerically the
incompressible three-dimensional MHD equations with a guide field
\cite{gomez_parallel_2005, gomez_mhd_2005}. All results reported here
are from runs with resolution of $N^3 = 512^3$ grid points. A
second-order Runge-Kutta time integration scheme is used. We use weak,
moderate and strong external magnetic fields, $B_0 = 0.25$, $1$ and
$8$ (in units of the initial r.m.s.~magnetic fluctuations value). We
also consider the case $B_0 = 0$ for reference with previous studies
\cite{servidio_time_2011}. Periodic boundary conditions are assumed in
all directions of a cube of side $2\pi L$ (where $L = 1$ is the
initial correlation length of the fluctuations, defined as the unit
length). Aliasing is removed by a two-thirds rule truncation
method. The initial state consists of nonzero amplitudes for the ${\bf
  v}({\bf k})$ and ${\bf b}({\bf k})$ fields, equipartioned in the
wave numbers within shells $1.1\leq k \leq 4$, with $k=|{\bf k}|$ (in
units of $2\pi L/\ell$ with $\ell$ the wavelength). Random phases have
been chosen for both fields. To achieve a statistically steady state
we consider a driving which consists of forcing terms added to
Eqs.~(\ref{eq:MHD_v}-\ref{eq:MHD_b}) in a fixed set of Fourier modes
in the band $0.9\leq k \leq 1.8$. The forcing has a random and a
time-coherent component, so that the correlation time of the forcing
is $\tau_f \approx 1$ (of the order of the unit time $t_0$).

The temporal range used to analyze the results is over $20$ unit times
for $B_0=0$ and $B_0=0.25$, over $25$ unit times for $B_0=1$, and over
$10$ unit times for $B_0=8$. All these time spans are considered after
the system reached a turbulent steady state, and we verified that they
were enough to ensure convergence of spectra and correlation
functions.

\section{Results}\label{sec_Res}

\subsection{Energy spectra and dominant time scales}

The axisymmetric energy spectrum $e(|\vec{k_\perp}| =
\sqrt{k_y^2+k_z^2}, k_\parallel = k_x, t)$, defined as
\begin{equation}\label{eq:axisymmetric}
\begin{split}  e(k_\perp, k_\parallel, t) = \sum_{\substack{k_\perp \leq |\vec{k}\times\hat{x}| < k_\perp+1 \\ k_\parallel \leq k_x < k_\parallel +1}} |\hat{u}(\vec{k},t)|^2 +|\hat{b}(\vec{k},t)|^2 = \\ = \int \left(|\hat{u}(\vec{k},t)|^2 +|\hat{b}(\vec{k},t)|^2\right) |\vec{k}| \sin \theta_k~d\phi_k,
\end{split}
\end{equation}
provides information on the anisotropy of the turbulence
relative to the the guide field \cite{mininni_isotropization_2012}. In
this study, the guide field is chosen along the $x$ axis, and thus the
wave vector components $k_\parallel$ and $k_\perp$, and the polar
angles in Fourier space $\theta_k$ and $\phi_k$, are relative to
this axis. In other words, in Eq.~(\ref{eq:axisymmetric}) 
$\theta_k = \arctan(k_\perp/k_\parallel)$ is the co-latitude in
Fourier space with respect to the axis with unit vector
$\hat{x}$ (that is, in the direction of the guide field), and
$\phi_k$ is the longitude with respect to the $y$-axis. The first
expression involving the summation in Eq.~(\ref{eq:axisymmetric}) is
the definition of the axisymmetric energy spectrum for a discrete
Fourier space (i.e., as used in the simulations), while the second
expression with the integral corresponds to the continuum limit. In
the following we treat both expressions as equivalent, replacing
integrals by summations when required for the numerics.

\begin{figure}
  \centering
  \includegraphics[width=1\columnwidth]{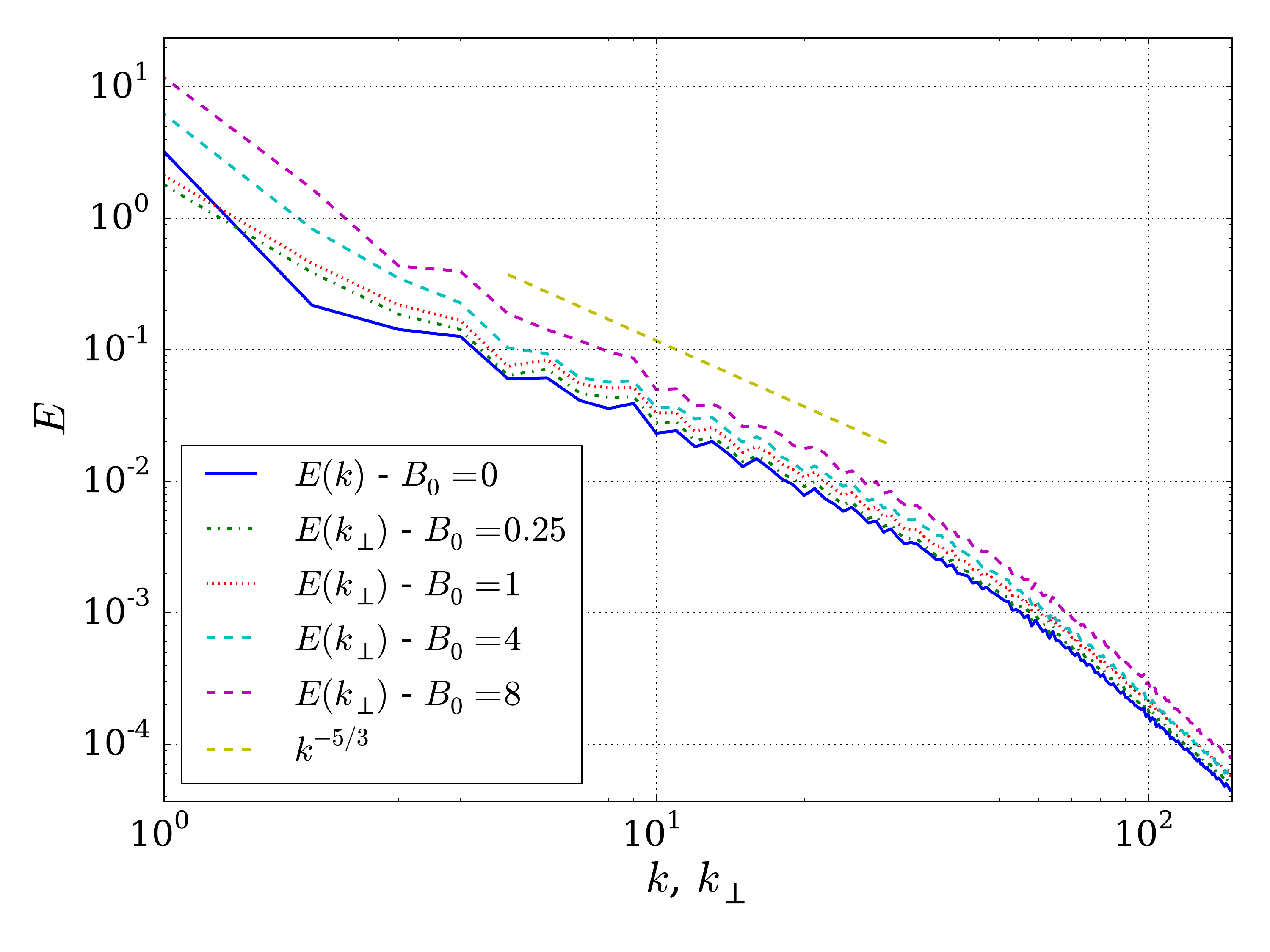}
  \caption{Reduced perpendicular energy spectra $E(k_\perp)$ for the
    simulations with $B_0=0.25$, $1$, $4$, and $8$, and isotropic energy
    spectrum $E(k)$ for the simulation with $B_0=0$. Kolmogorov
    scaling, $\sim k_\perp^{-5/3}$, is shown as reference.}
  \label{fig1:E}
\end{figure}

From the axisymmetric spectrum above, one can define the time averaged
reduced perpendicular energy spectrum $E(k_\perp)$
\cite{mininni_isotropization_2012} as
\begin{equation}\label{eq:reducedspectrum}
  E\left(k_\perp\right) = \frac{1}{T}\int\int e(|\vec{k_\perp}|,
  k_\parallel, t) \, dk_\parallel~dt,
\end{equation}
where we integrated over parallel wave numbers to obtain a spectrum
that depends only on $k_\perp$. Equivalently, the isotropic energy
spectrum $E(k)$ can be obtained from Eq.~(\ref{eq:axisymmetric}) by
integrating over $\theta_k$ in Fourier space. Figure \ref{fig1:E}
shows the isotropic energy spectrum $E(k)$ for the run with $B_0=0$,
and the reduced perpendicular energy spectrum $E(k_\perp)$ for the runs
with non-zero guide field.

\begin{figure*}
  \centering
%
%
  \subfigure[$B_0=0$]{\includegraphics[width=0.45\textwidth]{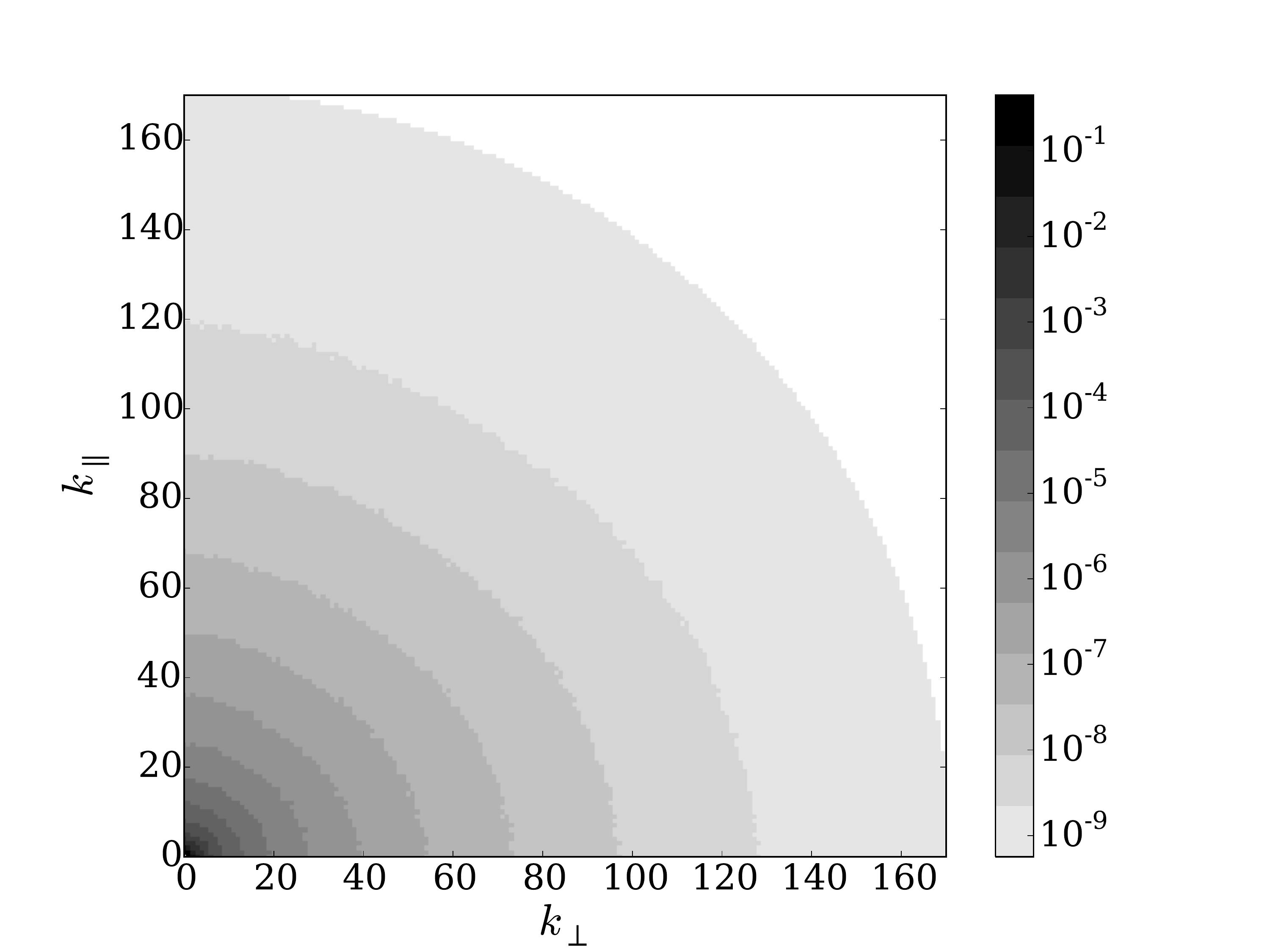}}
  \subfigure[$B_0=0.25$]{\includegraphics[width=0.45\textwidth]{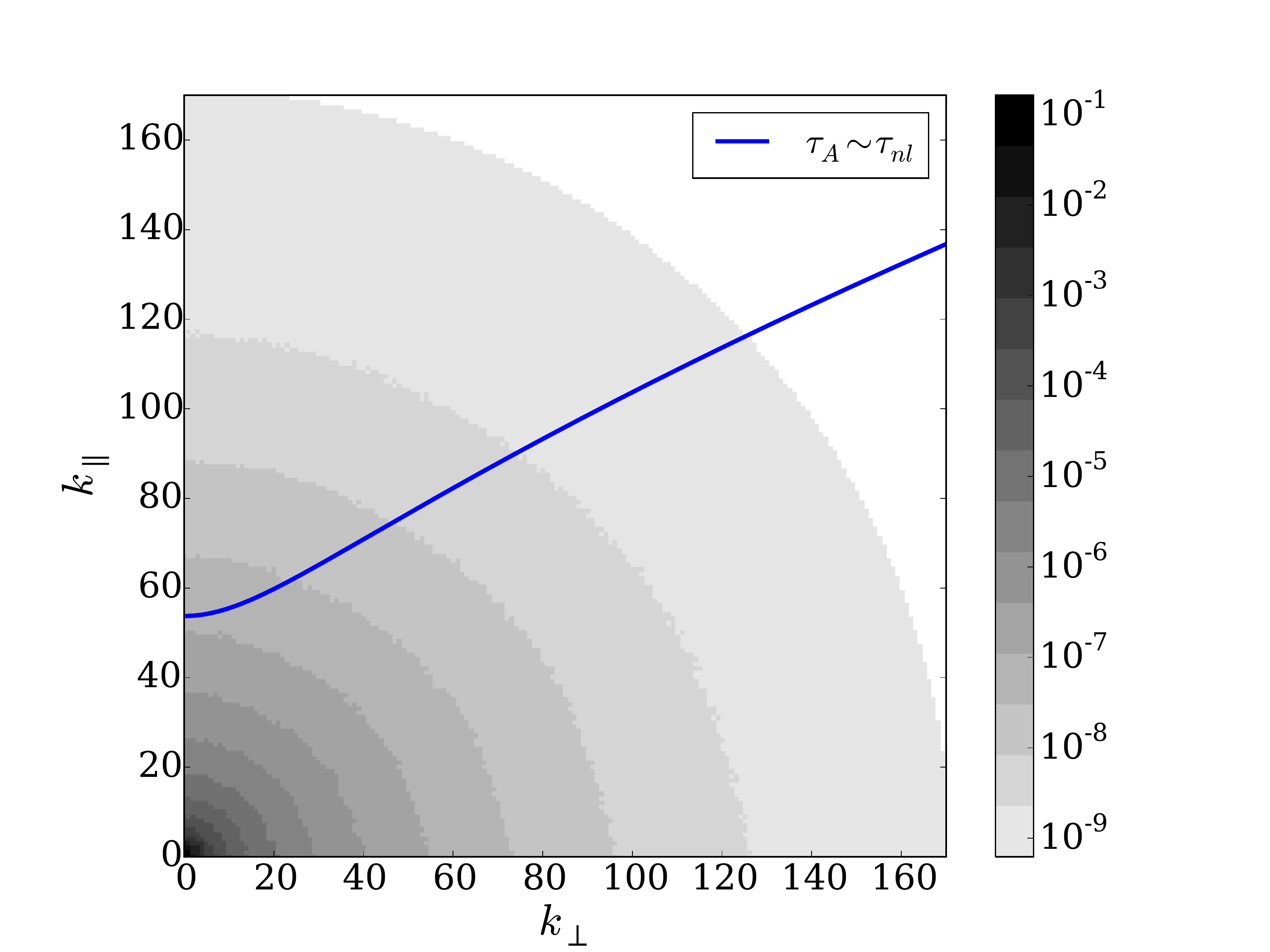}}

  \subfigure[$B_0=1$]{\includegraphics[width=0.45\textwidth]{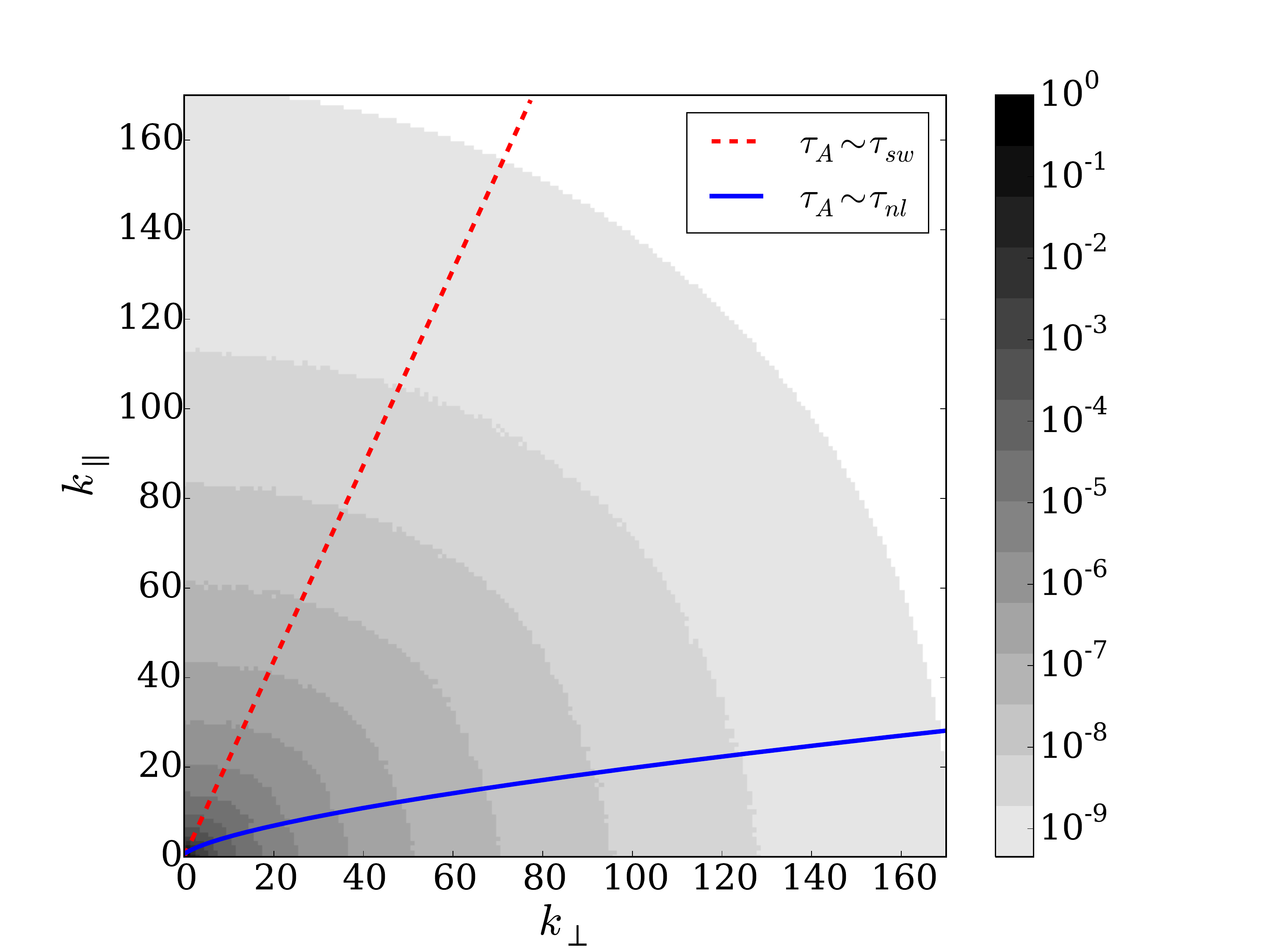}}
  \subfigure[$B_0=1$ with explanation]{\includegraphics[width=0.45\textwidth]{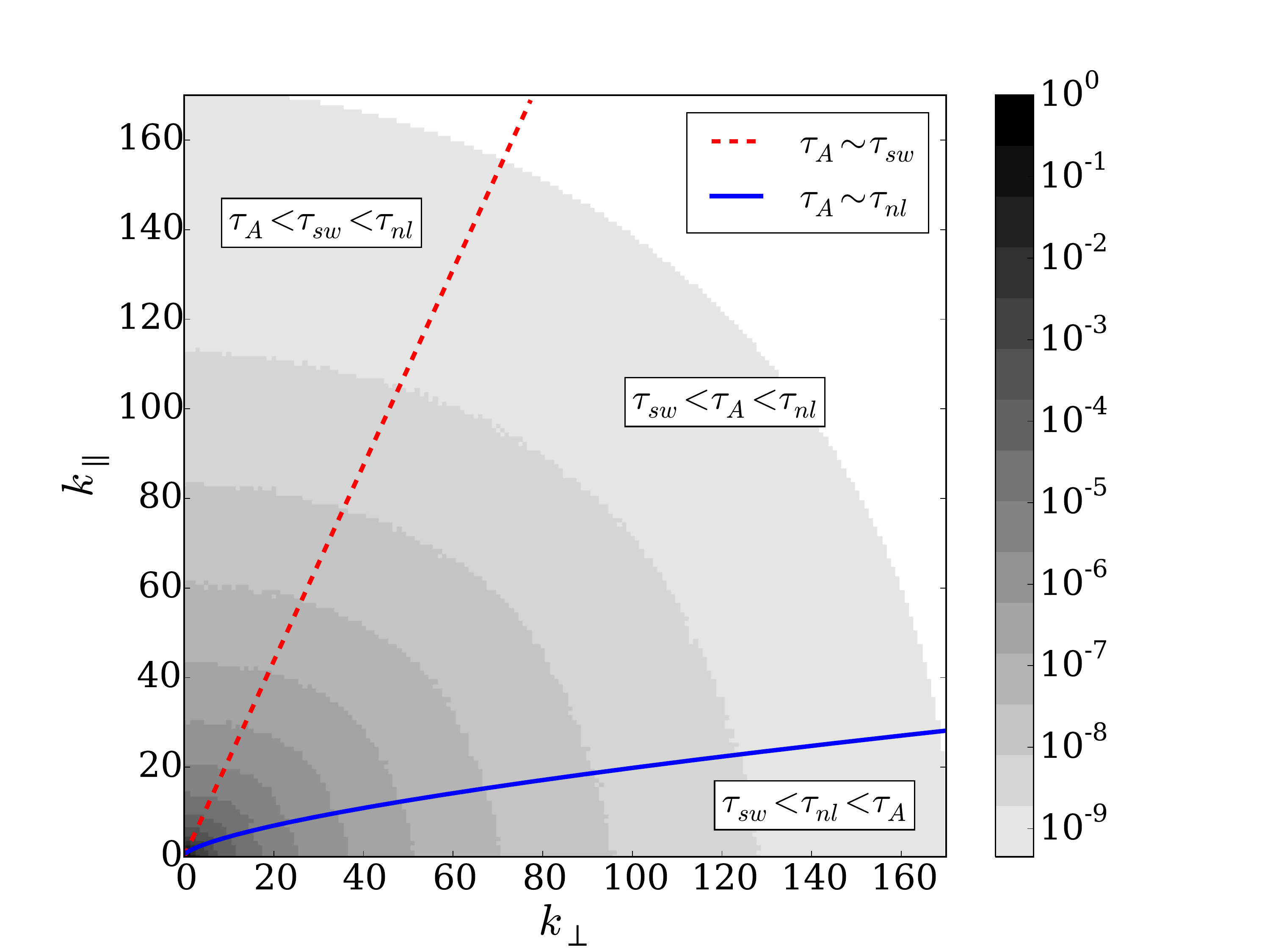}}

  \subfigure[$B_0=4$]{\includegraphics[width=0.45\textwidth]{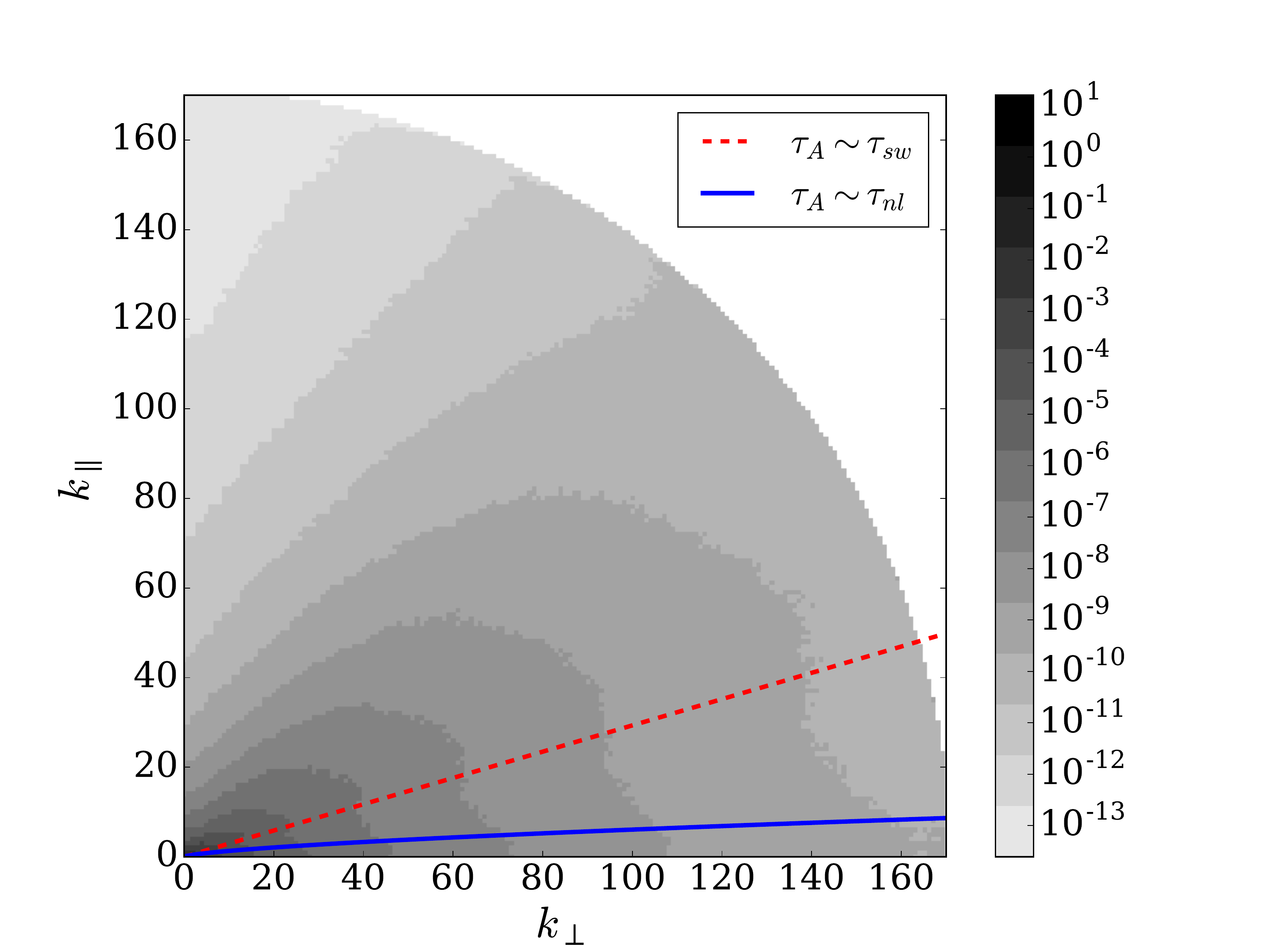}}
  \subfigure[$B_0=8$]{\includegraphics[width=0.45\textwidth]{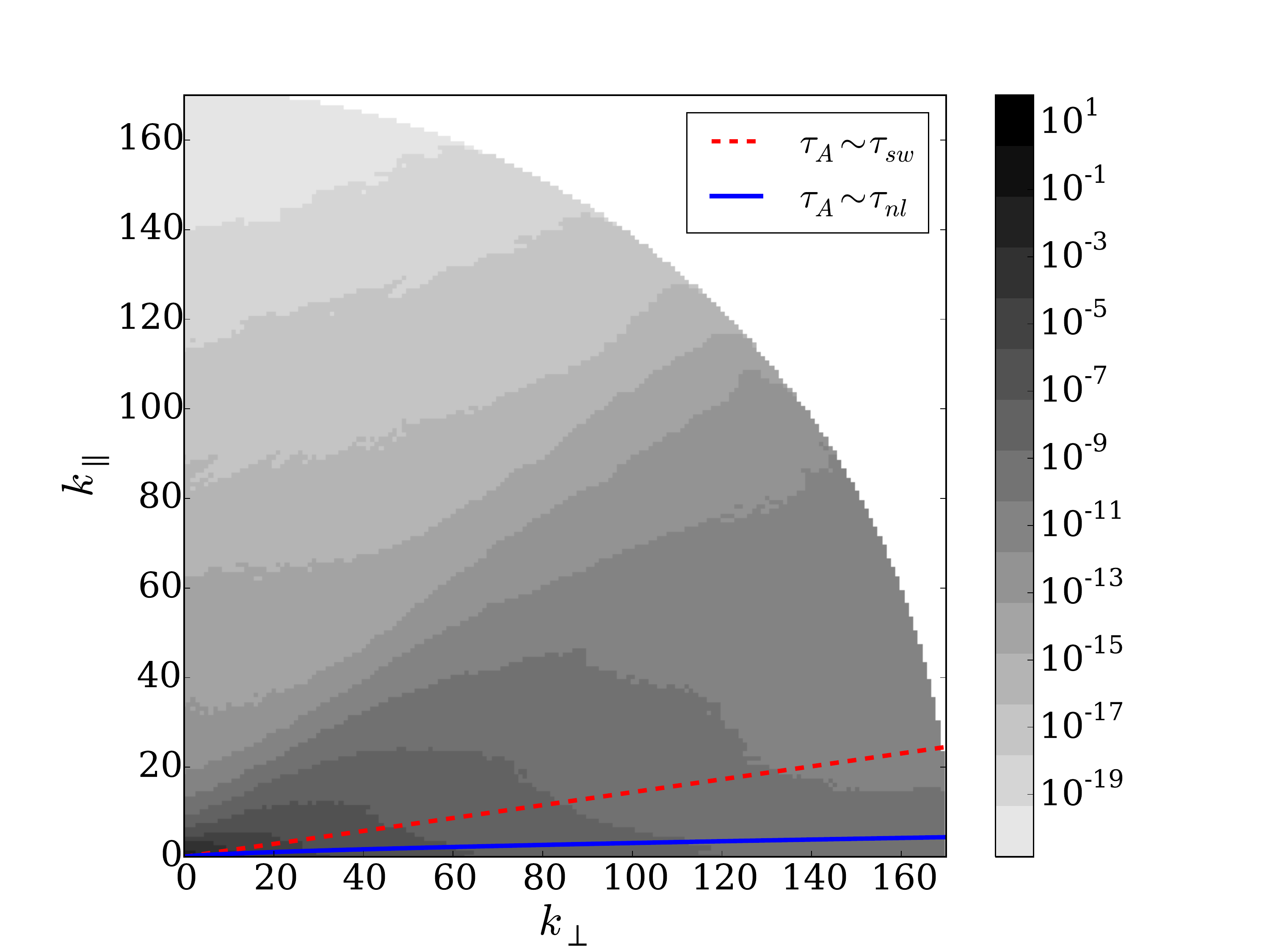}}
  \caption{Isocontours of the axisymmetric energy spectrum
    $e(k_\perp,k_\parallel)$ for different values of $B_0$. Dark
    means larger energy density (in logarithmic scale). The lines
    indicate the modes for which sweeping time or local non-linear
    time become equal to the Alfv\'en time. For large $B_0$ the
    isocontours change shape as they cross each of these lines. Note
    also the stronger anisotropy of the spectrum as $B_0$ increases,
    as well as the increase in the surface covered by modes which have
    the Alfv\'en period as the fastest time.}
  \label{fig2:isocontourns}
\end{figure*}

Figure \ref{fig2:isocontourns} shows contour plots of
$e(k_\perp,k_\parallel)/\sin(\theta_k)$, that is, the axisymmetric
spectrum (averaged in time), for the runs with $B_0=0$, $B_0=0.25$,
$B_0=1$, $B_0=4$, and $B_0=8$ respectively. For an isotropic flow
($B_0=0$, see Fig.~\ref{fig2:isocontourns}), contours of
$e(k_\perp,k_\parallel)/\sin(\theta_k)$ are circles as expected
\cite{mininni_isotropization_2012}. As the guide field intensity
increases, energy becomes more concentrated near the axis with
$k_\parallel=0$, evidencing the formation of elongated structures in
the direction of the guide field (or, in other words, of the relative
decrease of parallel gradients of the fields with respect to
perpendicular gradients).

The characteristic times defined in the Introduction, $\tau_A$,
$\tau_{sw}$, and $\tau_{nl}$, divide the Fourier space in
Fig.~\ref{fig2:isocontourns} in regions depending on how the
time scales are ordered:
\begin{equation}\label{eq:Asw} \tau_A < \tau_{sw} \hspace{0.05cm}
\Rightarrow \hspace{0.05cm} k_\perp <
\left(\sqrt{\left(\frac{B_0}{v_{rms}}\right)^2 \cdot
\left(\frac{C_{sw}}{C_A}\right)^2 - 1} \right) k_\parallel,
\end{equation}
\begin{equation}\label{eq:Anl} \tau_A < \tau_{nl} \hspace{0.05cm}
\Rightarrow \hspace{0.05cm} k_\perp <
\left(\sqrt{\left(\frac{B_0}{v_{rms}}\right)^3
\left(\frac{C_{nl}}{C_A}\right)^3 Lk_\parallel - 1} \right) 
k_\parallel,
\end{equation}
\begin{equation}\label{eq:nlsw} \tau_{nl} < \tau_{sw} \hspace{0.05cm}
\Rightarrow \hspace{0.05cm} \left( k_\perp^2 + k_\parallel^2
\right)^{1/6} < \frac{C_{sw}}{C_{nl}L^{1/3}}.
\end{equation}
In Figure \ref{fig2:isocontourns} we indicate the curves corresponding
to the modes that satisfy the relations $\tau_A\lesssim\tau_{sw}$ and
$\tau_A\lesssim\tau_{nl}$, respectively for $B_0=0$, $0.25$, $1$, $4$,
and $8$ (assuming, to plot all curves, that $C_{sw} \approx C_{nl}
\approx C_A \approx 1$; this choice will be later confirmed by the
analysis of the correlation functions).

As we can see from Eq.~(\ref{eq:nlsw}), the region where
$\tau_{nl}\leq\tau_{sw}$ is a small circle around the origin, where
$k_\perp^2 + k_\parallel^2 \leq (C_{sw}/L^{1/3}C_{nl})^6 \approx 1$, and
is not shown in the figure. Modes outside the region with
$\tau_{nl}<\tau_{sw}$ should decorrelate with the sweeping time or the
Alfv\'en time, depending on which one is fastest. Equation 
(\ref{eq:Asw}) tells us that in the area to the left of the curve
$\tau_A\sim\tau_{sw}$ we have $\tau_A<\tau_{sw}$, while 
Eq.~(\ref{eq:Anl}) tells us that in the area to the left of the curve
$\tau_A\sim\tau_{nl}$ we have $\tau_A<\tau_{nl}$ (see
Fig.~\ref{fig2:isocontourns}d). For the largest value of $B_0$
considered (i.e., the simulation with $B_0=8$), most of the modes have
the Alfv\'en period as the fastest time (i.e., the largest area in the
plot is above and to the left of the curve $\tau_A \sim \tau_{sw}$),
although a significant fraction of the energy in the system is not in
these modes as it concentrates instead near the axis with 
$k_\parallel = 0$.

\subsection{Spatio-temporal spectra}

\begin{figure}
  \centering 
  \includegraphics[width=1\columnwidth]{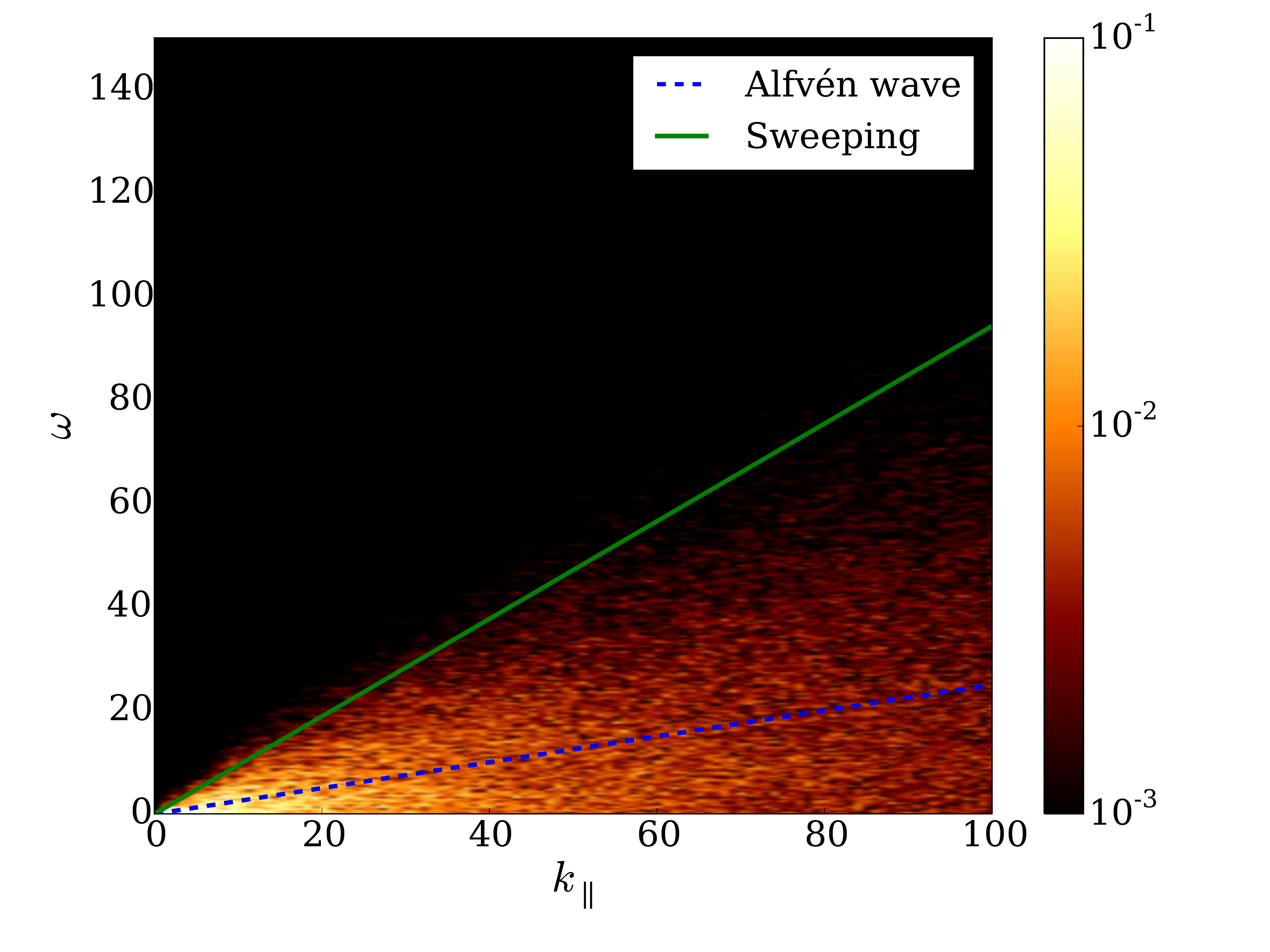}
  \caption{Normalized wave vector and frequency spectrum $E({\bf k},
    \omega)/E({\bf k})$ for the run with ${\bf B_0}=0.25$, for modes
    with $k_\perp=0$, and thus as a function of $k_\parallel$. Lighter
    regions indicate larger energy density. The spectrum corresponds
    to the power in the time and space Fourier transform of the
    fields, such that accumulation of energy in modes near the
    dispersion relation (or in all modes below the sweeping curve)
    indicates dominance of a physical effect (i.e., of its associated
    frequency) in the dynamics of a given scale $\sim
    1/k_\parallel$. The dashed (blue) line indicates the dispersion
    relation for Alfv\'en waves, and the continuous (green) line
    indicates the sweeping relation. A broad excitation of modes is
    observed for all modes with $\omega \leq v_{rms} k_\parallel$
    (sweeping), while only a very weak accumulation at small
    $k_\parallel$ can be seen for $\omega=B_0 k_\parallel$
    (Alfv\'en).}
  \label{fig3:B025_bvf_Etot_kperp0}
\end{figure}

\begin{figure}
  \centering
  \includegraphics[width=1\columnwidth]{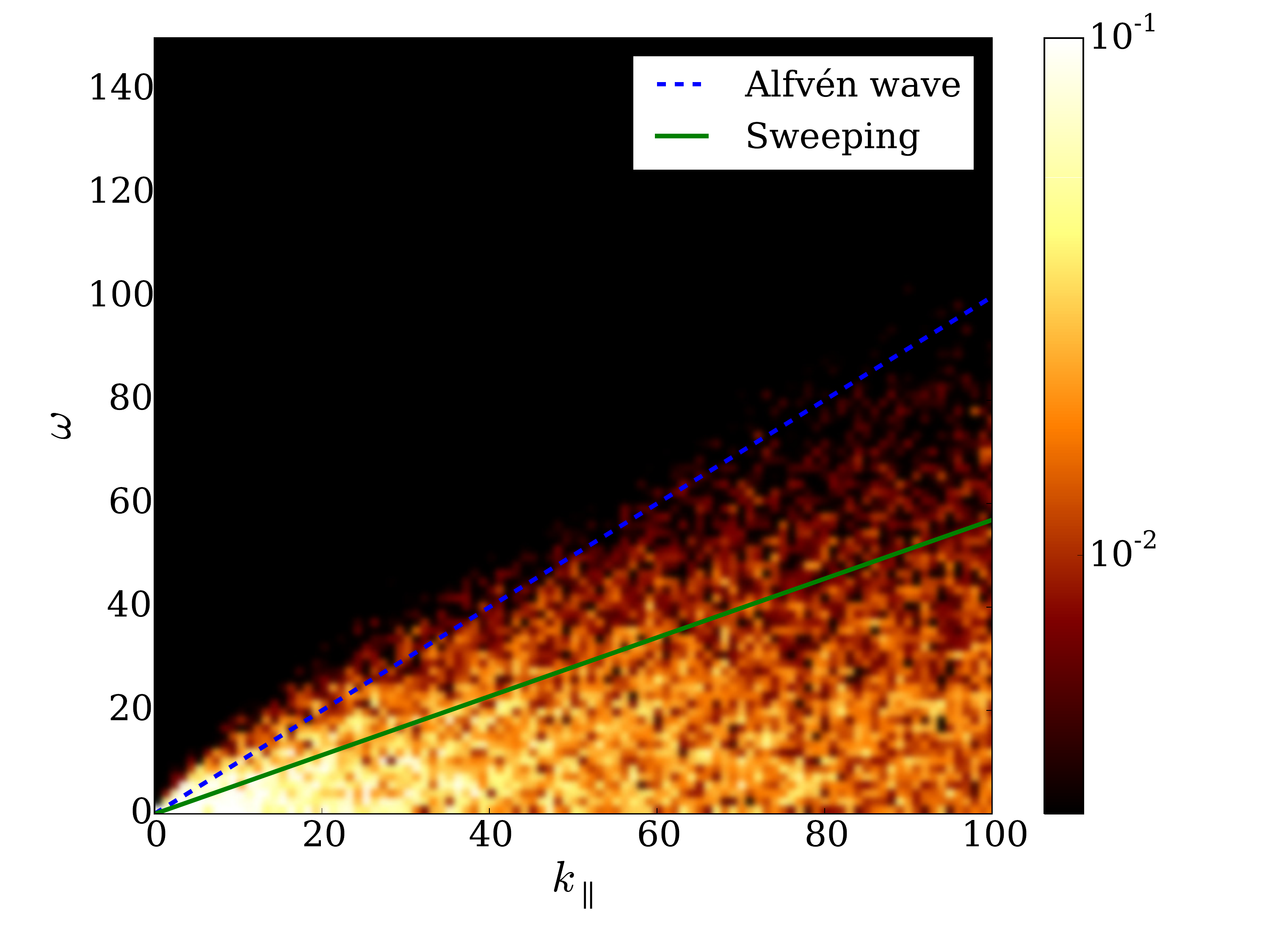}
  \caption{Normalized wave vector and frequency spectrum $E({\bf k},
    \omega)/E({\bf k})$ for the run with $B_0=1$, for modes with
    $k_\perp=0$, and thus as a function of $k_\parallel$ and
    $\omega$. Lighter regions indicate larger energy density. The
    dashed (blue) line indicates the dispersion relation for Alfv\'en
    waves and the continuous (green) line indicates the sweeping
    relation.}
  \label{fig3:B1_bvf_Etot_kperp0}
\end{figure}

\begin{figure}
  \centering
  \includegraphics[width=1\columnwidth]{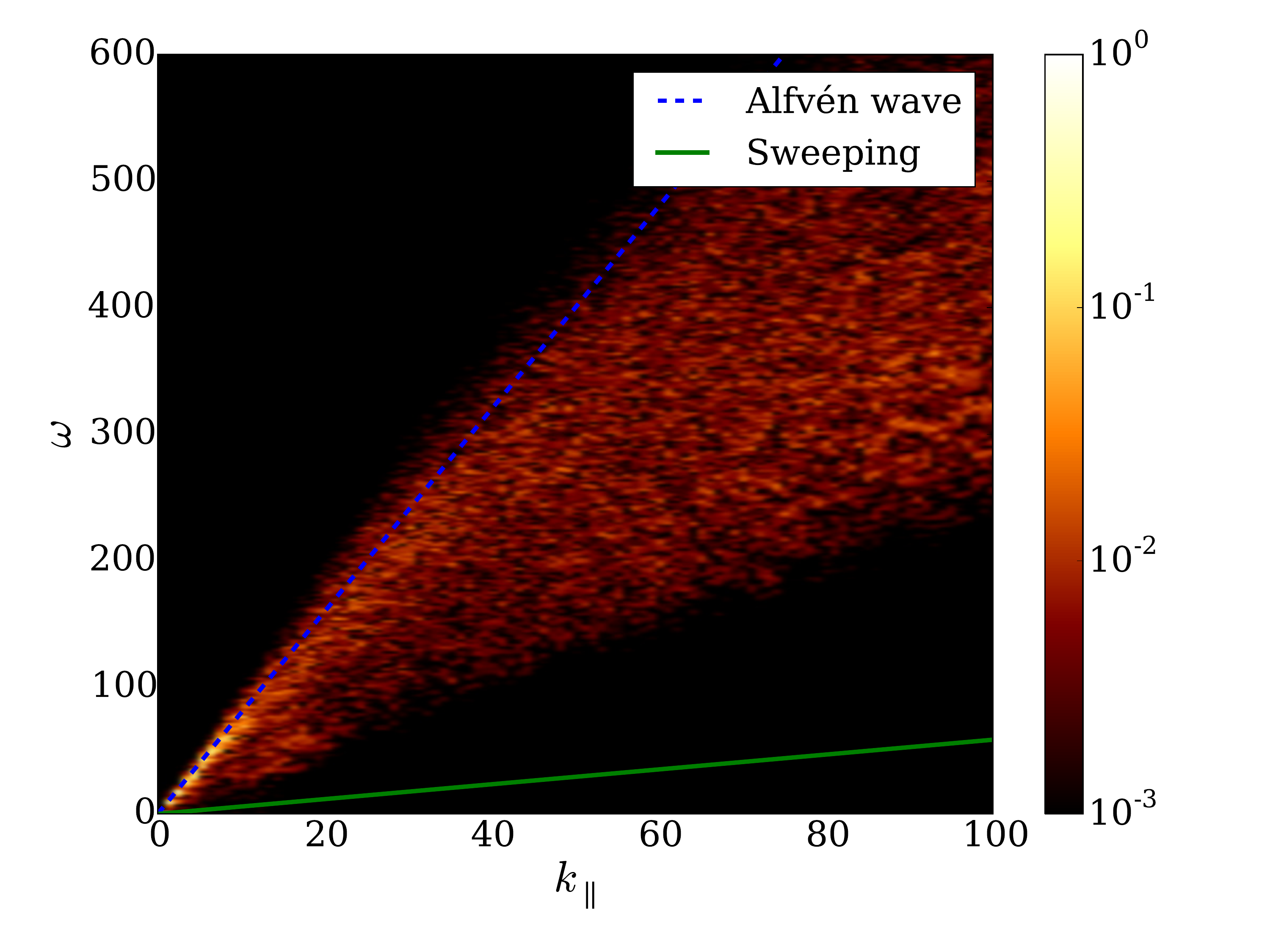}
  \caption{Normalized wave vector and frequency spectrum $E({\bf k},
    \omega)/E({\bf k})$ for the run with $B_0=8$, for modes with
    $k_\perp=0$, and thus as a function of $k_\parallel$ and
    $\omega$. Lighter regions indicate larger energy density. The
    dashed (blue) line indicates the dispersion relation for Alfv\'en
    waves and the continuous (green) line indicates the sweeping
    relation. Note in this case power is concentrated in a narrow
    region near the wave dispersion relation up to 
    $k_\parallel \approx 10$, corresponding to Alfv\'enic
    excitations.}
  \label{fig3:B8_bvf_Etot_kperp0}
\end{figure}

Figure \ref{fig3:B025_bvf_Etot_kperp0} (for the simulation with
$B_0=0.25$), Fig.~\ref{fig3:B1_bvf_Etot_kperp0} ($B_0=1$), and
Fig.~\ref{fig3:B8_bvf_Etot_kperp0} ($B_0=8$) show the wave vector and
frequency spectrum $E(\vec{k}, \omega)/E(\vec{k})$ for modes $\vec{k}$ 
with $k_\perp = 0$, where
\begin{equation}
  E(\vec{k})=\int E(\vec{k},\omega)d\omega
\end{equation}
is the total energy spectrum. With this choice for the
normalization, the frequencies that concentrate most of the energy for
each $\vec{k}$ are more clearly visible. For $B_0=0.25$ 
(Fig.~\ref{fig3:B025_bvf_Etot_kperp0}) we observe a spread of the
energy concentration clearly below the sweeping relation line (i.e.,
we see excitations in all modes with frequency equal or smaller than
$\omega = v_{rms} k_\parallel$, indicating small scale structures are
advected by all velocities equal and smaller than $v_{rms}$). A weak
accumulation near the Alfv\'enic dispersion relation $\omega = B_0
k_\parallel$ is also visible for small wavenumbers $k_\parallel$,
although the broad spectrum (in the frequency domain) suggests
sweeping is dominant in this case.

As the mean field increases to $B_0=1$ 
(Fig.~\ref{fig3:B1_bvf_Etot_kperp0}), some of the energy is
concentrated above the sweeping line and starts to follow the
Alfv\'enic dispersion relation, although the spectrum is still broad
in frequencies, with a large fraction of the energy below the sweeping
relation. This behavior changes drastically for larger values of
$B_0$. In Figure \ref{fig3:B8_bvf_Etot_kperp0} ($B_0=8$) we can see
energy clearly concentrating around the dispersion relation of Alfv\'en
waves, with the power sharply peaked around the wave modes up to
$k_\parallel \approx 10$, and then suddenly broadening towards the
sweeping relation for larger wavenumbers. Note that this indicates a
competition between the magnetohydrodynamic sweeping time and the
Alfv\'en time, with the former becoming dominant at large scales for
large values of $B_0$. These results support and enhance the ones
obtained by Dmitruk and Matthaeus \cite{dmitruk_waves_2009}, and are
compatible for small wavenumber and large $B_0$ with those recently
obtained in \cite{meyrand_direct_2016, meyrand_weak_2015}. In
particular, \cite{meyrand_direct_2016} also reported a transition from
a narrow wave spectrum to a broader spectrum, although the scale and
mechanism responsible for the transition was not clear. As will be
confirmed next from the correlation functions, the competition
between sweeping and the Alfv\'en time as the dominant decorrelation
time is responsible for the change observed in the behavior of the
spectrum.

\begin{figure}
  \centering
  \subfigure[$\Gamma(k_\perp=0,k_\parallel=k_0,\tau)$]{\includegraphics[width=1\columnwidth]{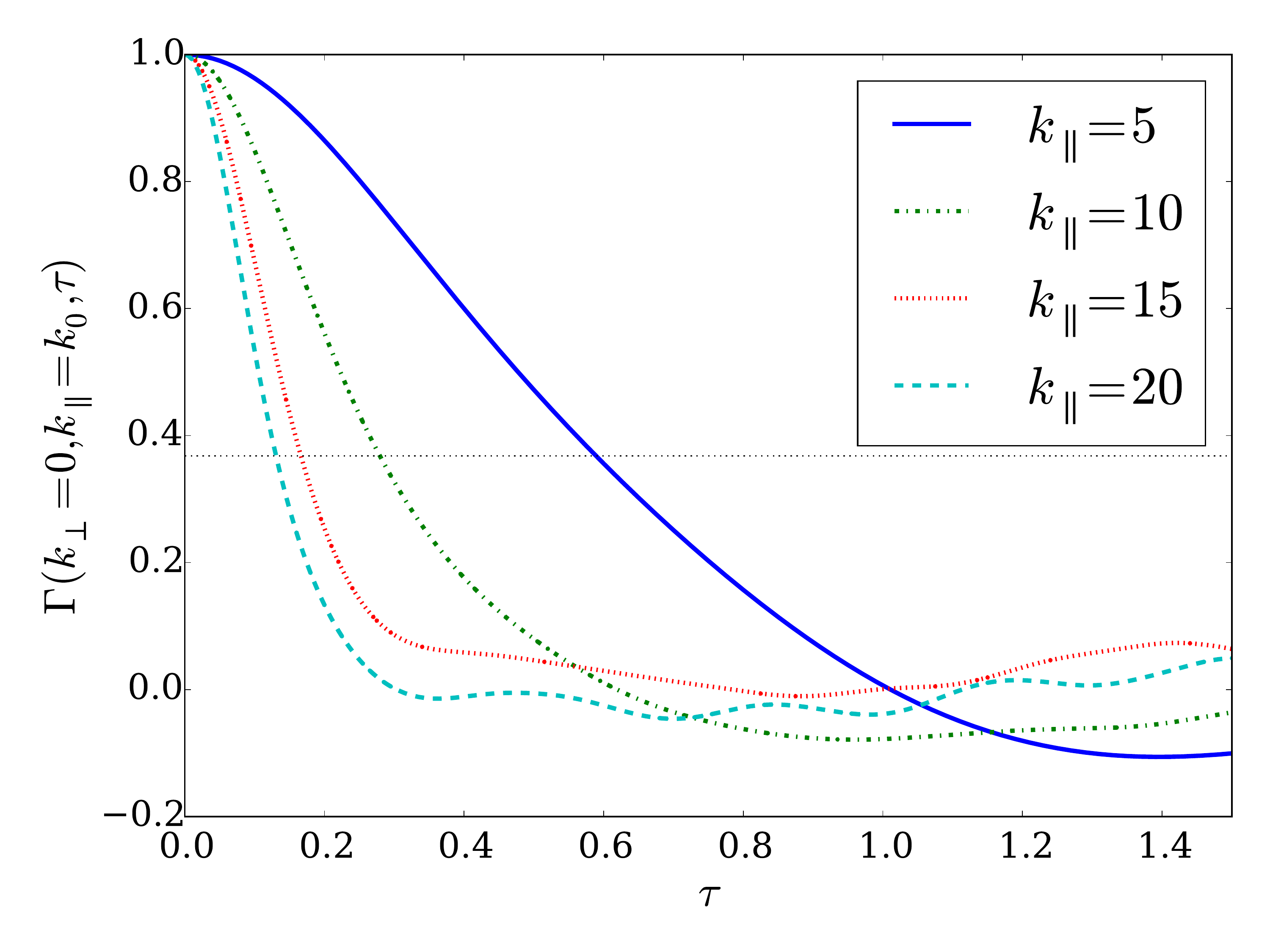}}

  \subfigure[$\Gamma(k_\perp=k_0,k_\parallel=0,\tau)$]{\includegraphics[width=1\columnwidth]{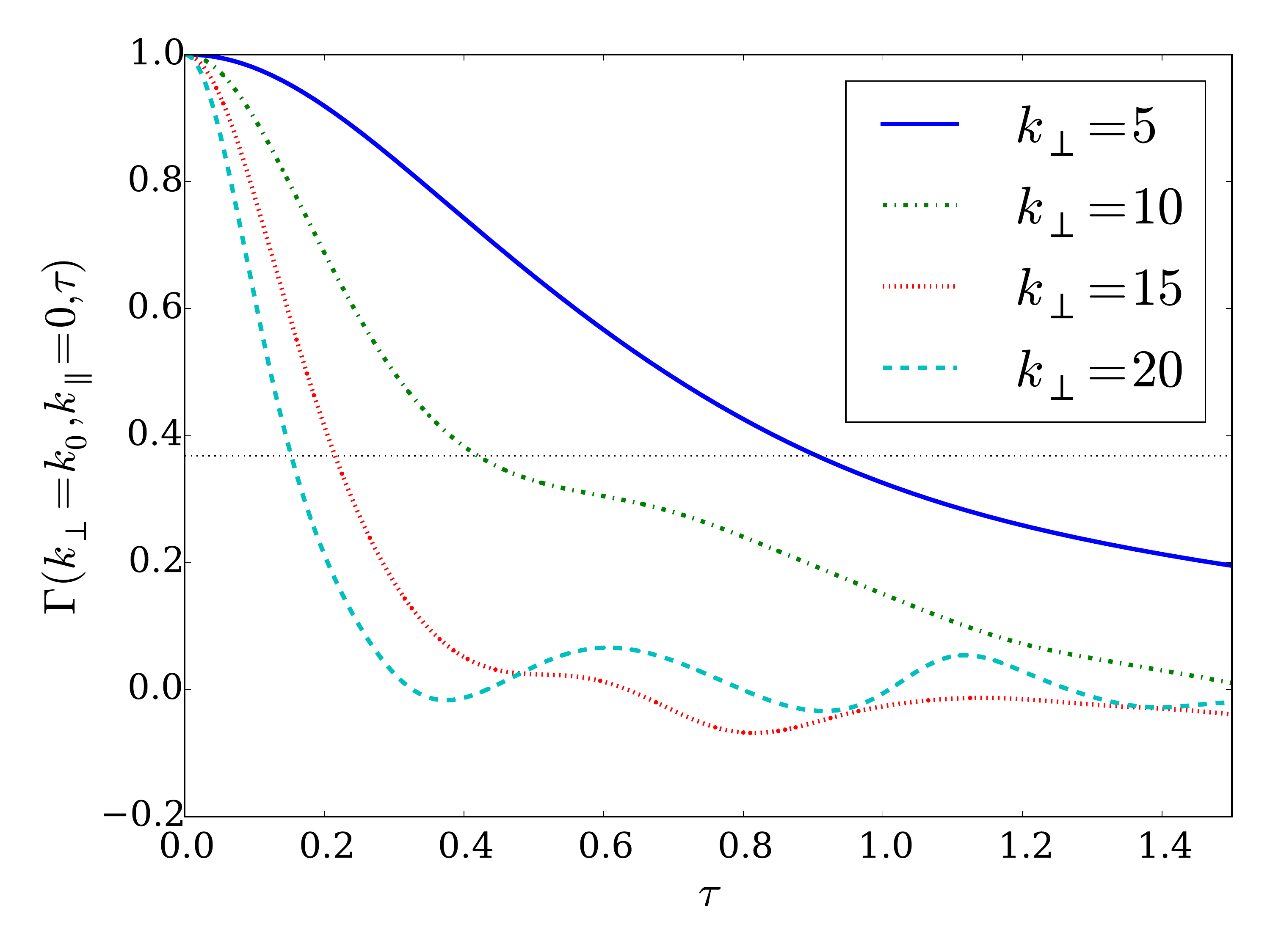}}
  \caption{Correlation functions
    $\Gamma(k_\perp=0,k_\parallel=k_0,\tau)$ and
    $\Gamma(k_\perp=k_0,k_\parallel=0,\tau)$ as a function of the lag
    time $\tau$, for $k_0=5$, 10, 15, and 20, in the simulation with
    $B_0=1$. The value of $\tau$ for which $\Gamma=1/e$ (horizontal
    dotted line) corresponds to the decorrelation time $\tau_D$ for
    each value of $\vec{k}$.}
  \label{fig4:B1_bvf_b_kperp/kpara0}
\end{figure}

\subsection{Correlation functions and decorrelation times}

In order to discern between the different phenomena (and relevant time
scales) acting in magnetohydrodynamic turbulence, we studied the
correlation functions $\Gamma(\vec{k},\tau)$ as explained in detail
before in Sec.~\ref{sec_Wfspectrum_and_Gamma}. Since we focus on
turbulence with a guide magnetic field, we use
$\Gamma(k_\perp,k_\parallel,\tau)$ and consider several values of
$(k_\perp,k_\parallel)$ to study the decorrelation as a function of
the time lag $\tau$ at different scales. In 
Fig.~\ref{fig4:B1_bvf_b_kperp/kpara0}, the correlation functions
$\Gamma(k_\perp=0,k_\parallel=k_0,\tau)$ and
$\Gamma(k_\perp=k_0,k_\parallel=0,\tau)$ are shown for different
values of $k_0$ for the moderate external magnetic field $B_0=1$. Here
we can see the typical behavior of correlation functions, with the
largest scales (smallest $k$) taking longer time to
decorrelate. Similar results were found for the other external
magnetic field considered, $B_0=0$, $0.25$, $4$, and $8$.

To understand which of the different times (non-linear time, random
sweeping, and Alfv\'en propagation) are controlling the temporal
decorrelation, we need to compare the scaling of the decorrelation
time with the theoretical scale dependence expected for each physical
process. In order to do this, we use the fact that the mode with wave 
vector $\vec{k}$ should be decorrelated after a time $\tau_D(\vec{k})$
following an approximate exponential decay
\begin{equation}
\Gamma(\vec{k},\tau) \sim e^{-\tau/\tau_D(\vec{k})}.
\end{equation}
For simplicity, we will evaluate $\tau_D(\vec{k})$ as the time at which
the function $\Gamma$ decays to $1/e$ of its initial value.

\begin{figure}
  \includegraphics[width=1\columnwidth]{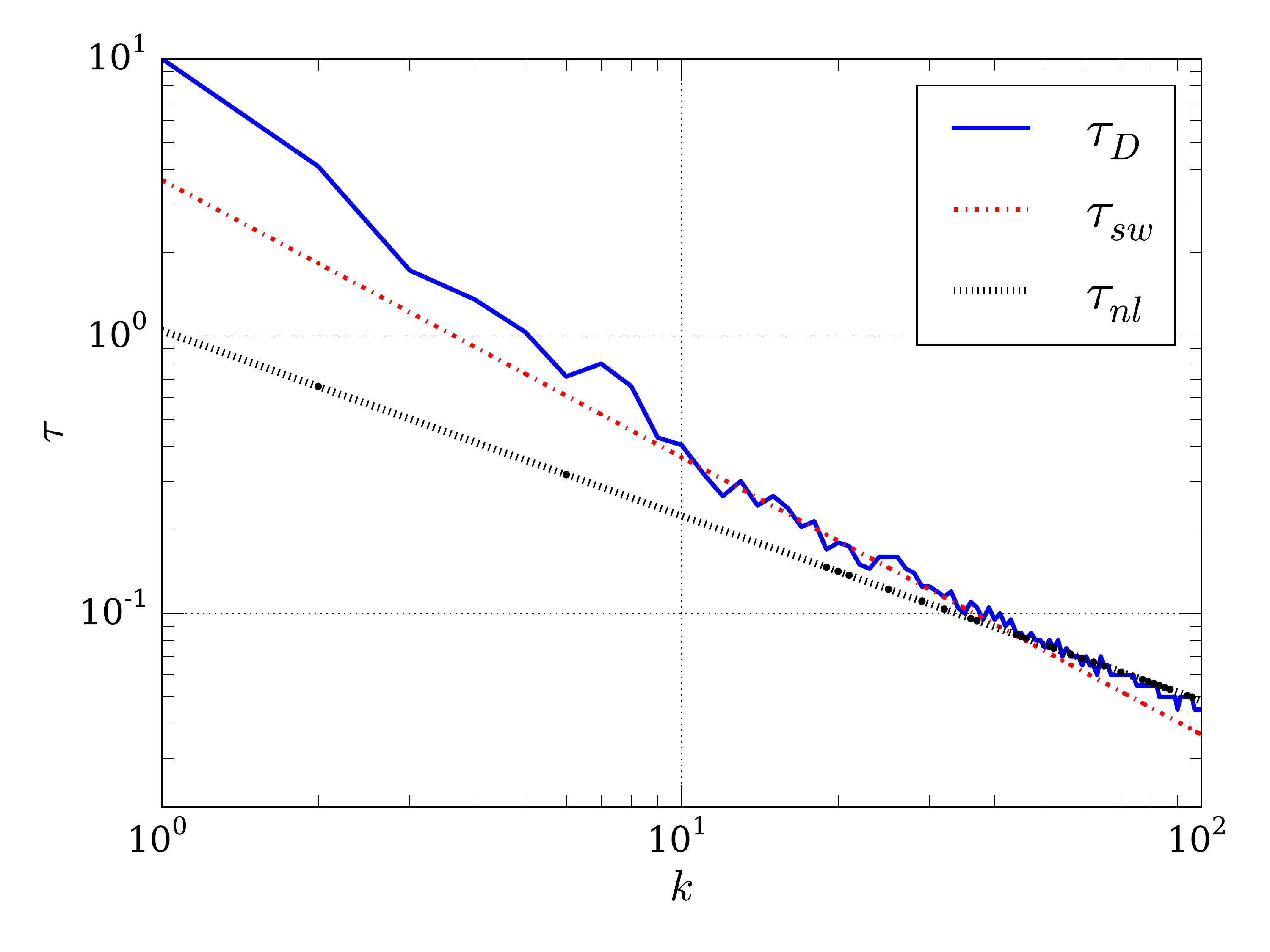}
  \caption{Decorrelation times as a function of $k=|\vec{k}|$ for the
    isotropic case $B_0=0$. The straight lines indicate the
    theoretical predictions corresponding to the sweeping time and the
    nonlinear time. Except at the largest wavenumbers, the
    decorrelation time seems to be dominated by sweeping.}
  \label{fig5:B0_bvf_b_kpara_0}
\end{figure}

As a first example, Fig.~\ref{fig5:B0_bvf_b_kpara_0} shows the
decorrelation time $\tau_D$ obtained from $\Gamma(k,\tau)$ in the
isotropic case with $B_0=0$. We can see that the decorrelation time
scales in good agreement with the sweeping time, except perhaps at the
largest wavenumbers (smallest scales). These results are consistent
with the ones obtained by Servidio {\it et al}
\cite{servidio_time_2011} in the isotropic case.

\begin{figure}
  \centering
  \subfigure[$k_\perp=0$]{\includegraphics[width=1\columnwidth]{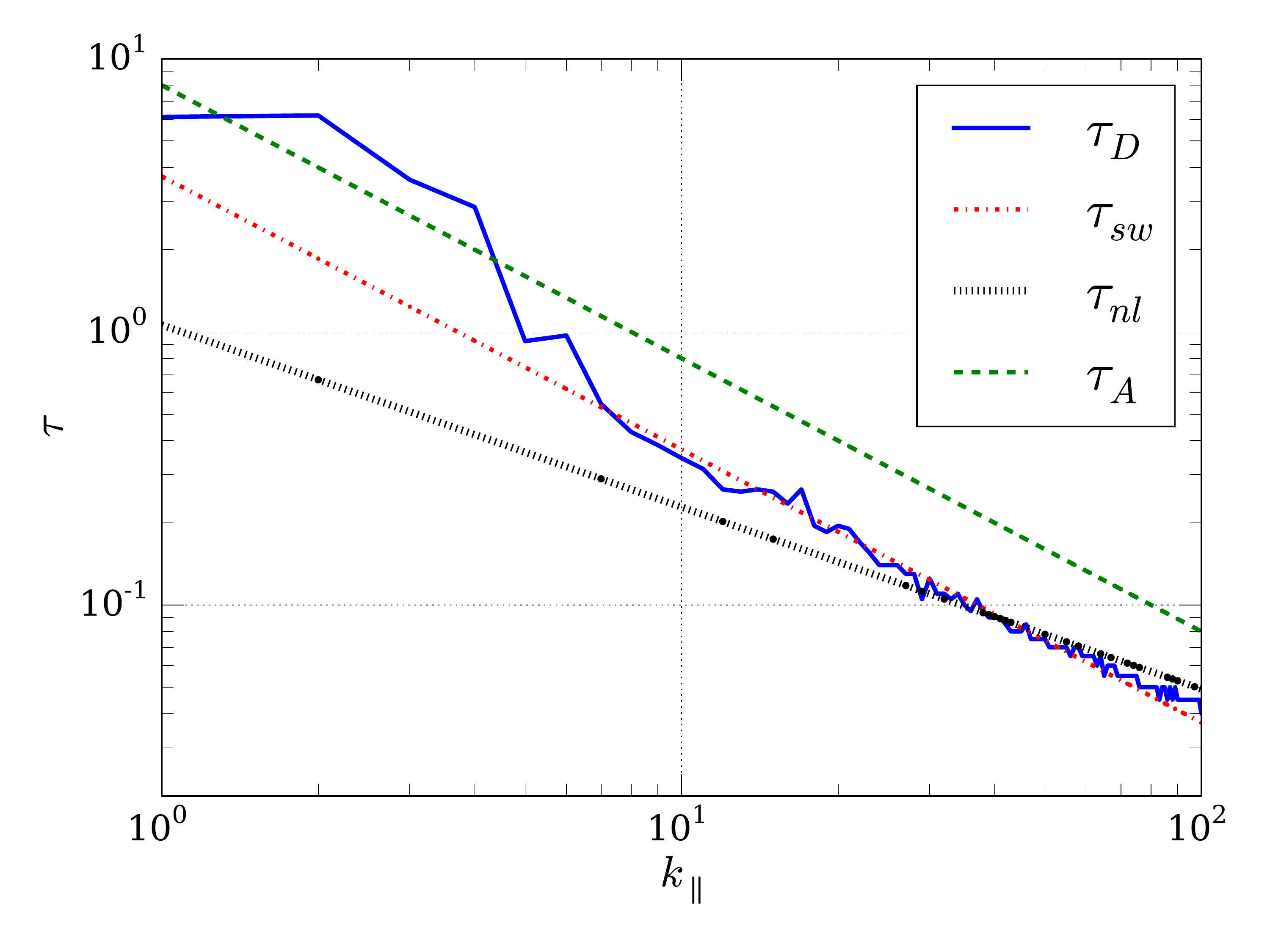}}

  \subfigure[$k_\perp=10$]{\includegraphics[width=1\columnwidth]{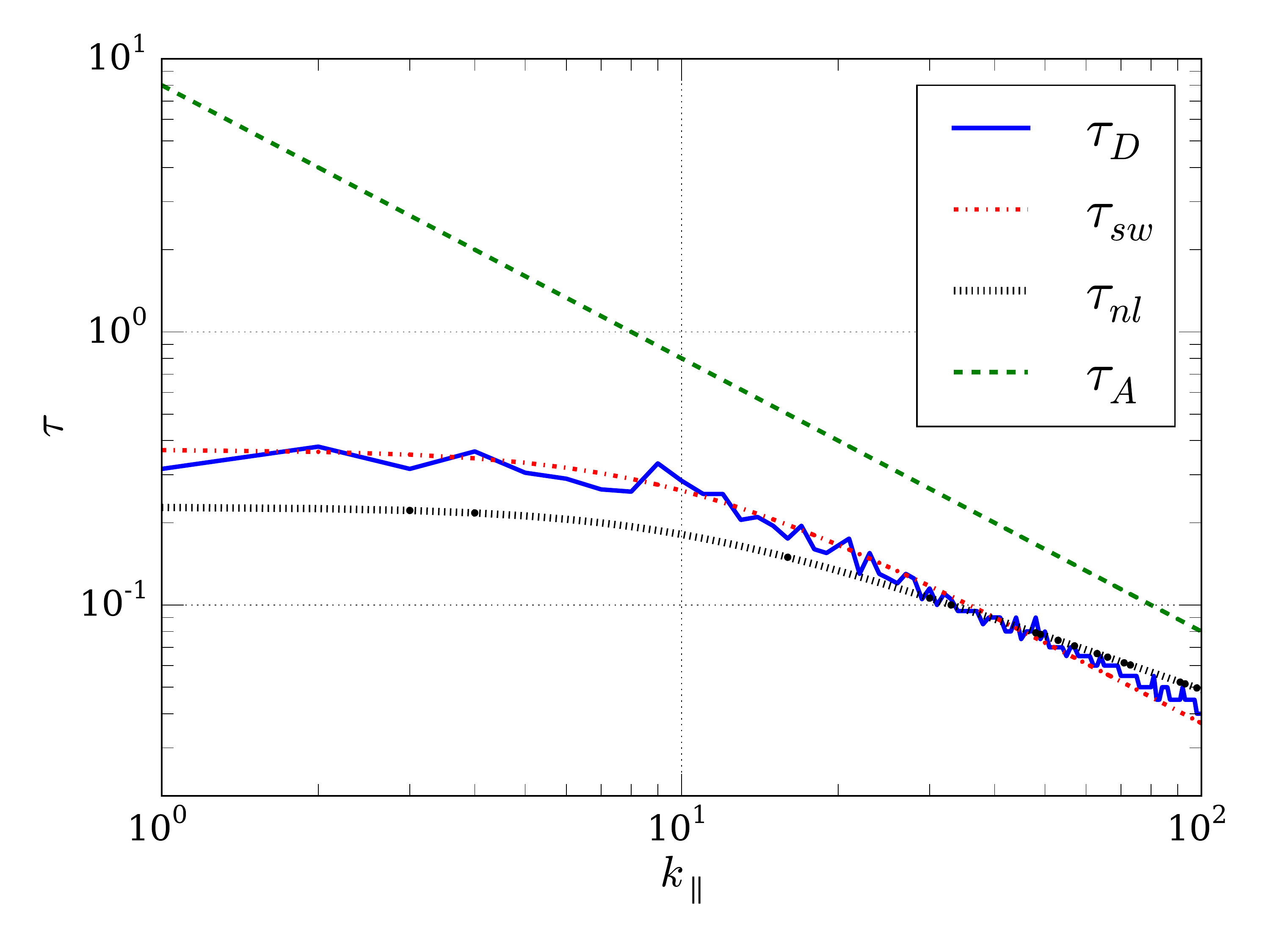}}

  \subfigure[$k_\perp=20$]{\includegraphics[width=1\columnwidth]{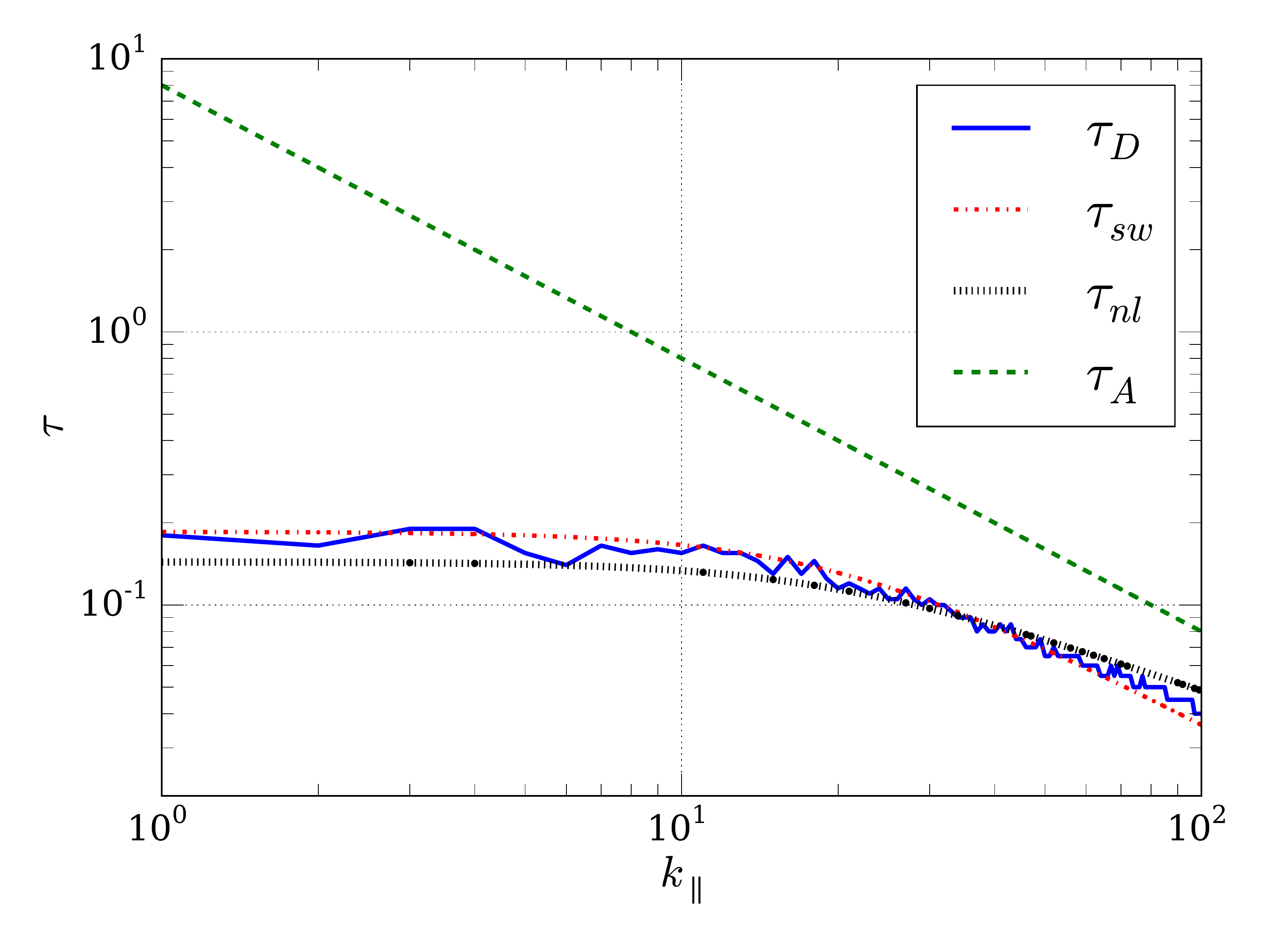}}
  \caption{Decorrelation times $\tau_D$ for the run with 
    $B_0=0.25$. In each panel $k_\perp$ is held constant and $k_\parallel$
    is varied; (a) $k_\perp = 0$, (b) $k_\perp = 10$, and (c) $k_\perp =
    20$. The lines indicate theoretical predictions for the scaling of
    several physical time scales. The measured value of $\tau_D$ is always
    close to $\tau_{sw}$, except for $k_\perp = 0$ and $k_\parallel$
    between $\approx 1$ and 5 for which the dominant time scale is the
    Alfv\'en time.}
  \label{fig5:B025_bvf_b_kperp}
\end{figure}

\begin{figure}
  \centering
  \subfigure[$k_\parallel=0$]{\includegraphics[width=1\columnwidth]{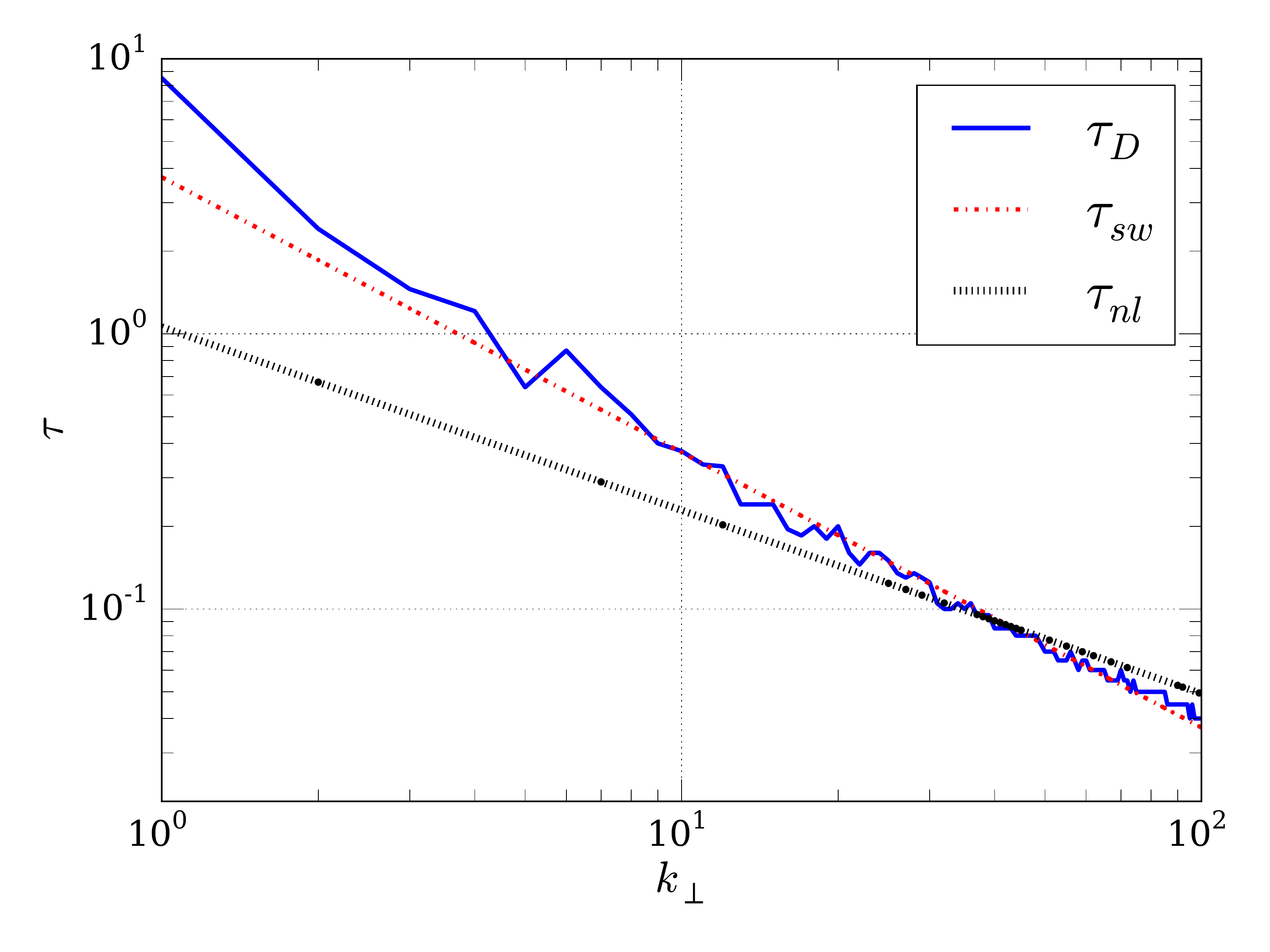}}

  \subfigure[$k_\parallel=10$]{\includegraphics[width=1\columnwidth]{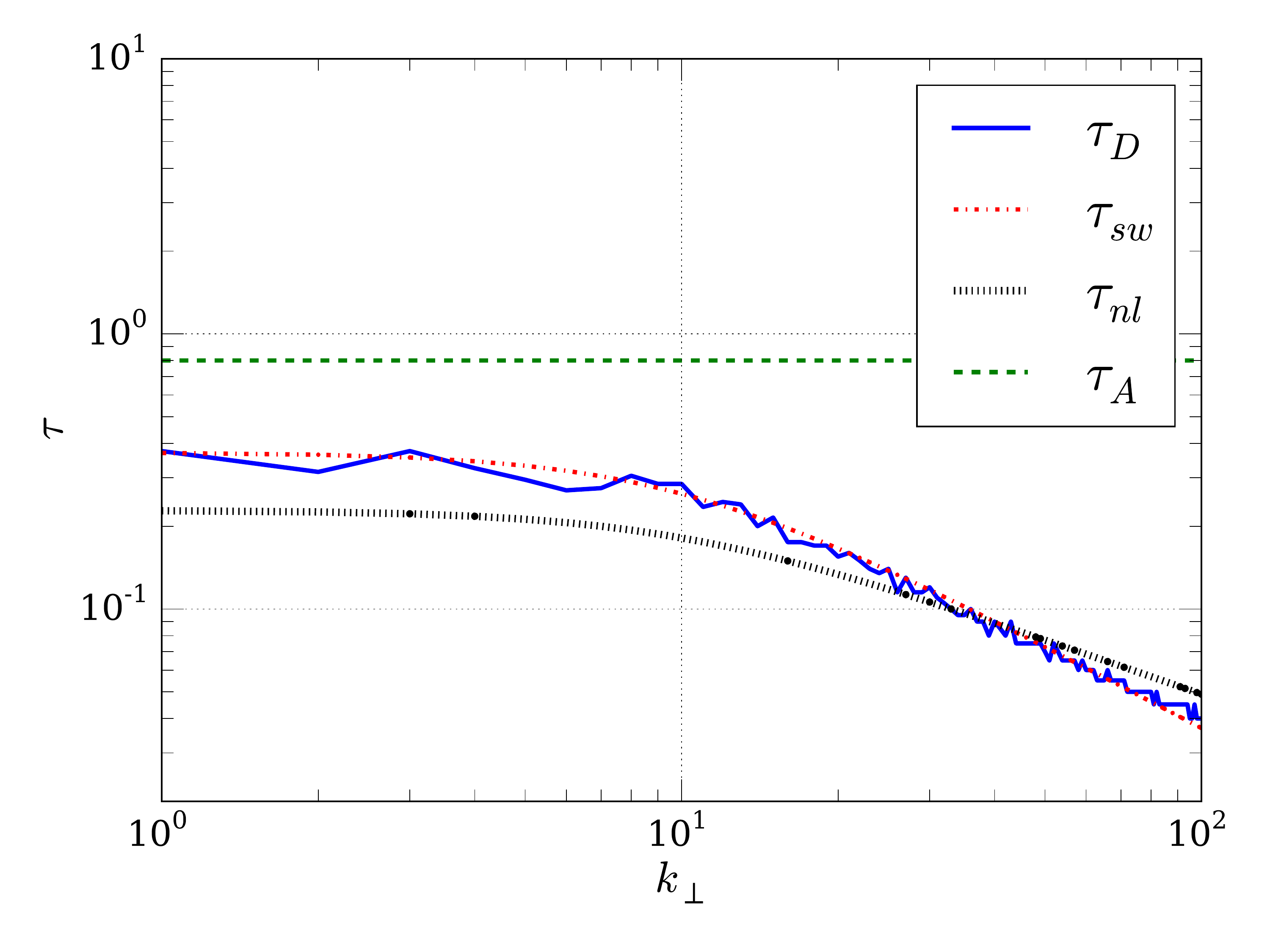}}

  \subfigure[$k_\parallel=20$]{\includegraphics[width=1\columnwidth]{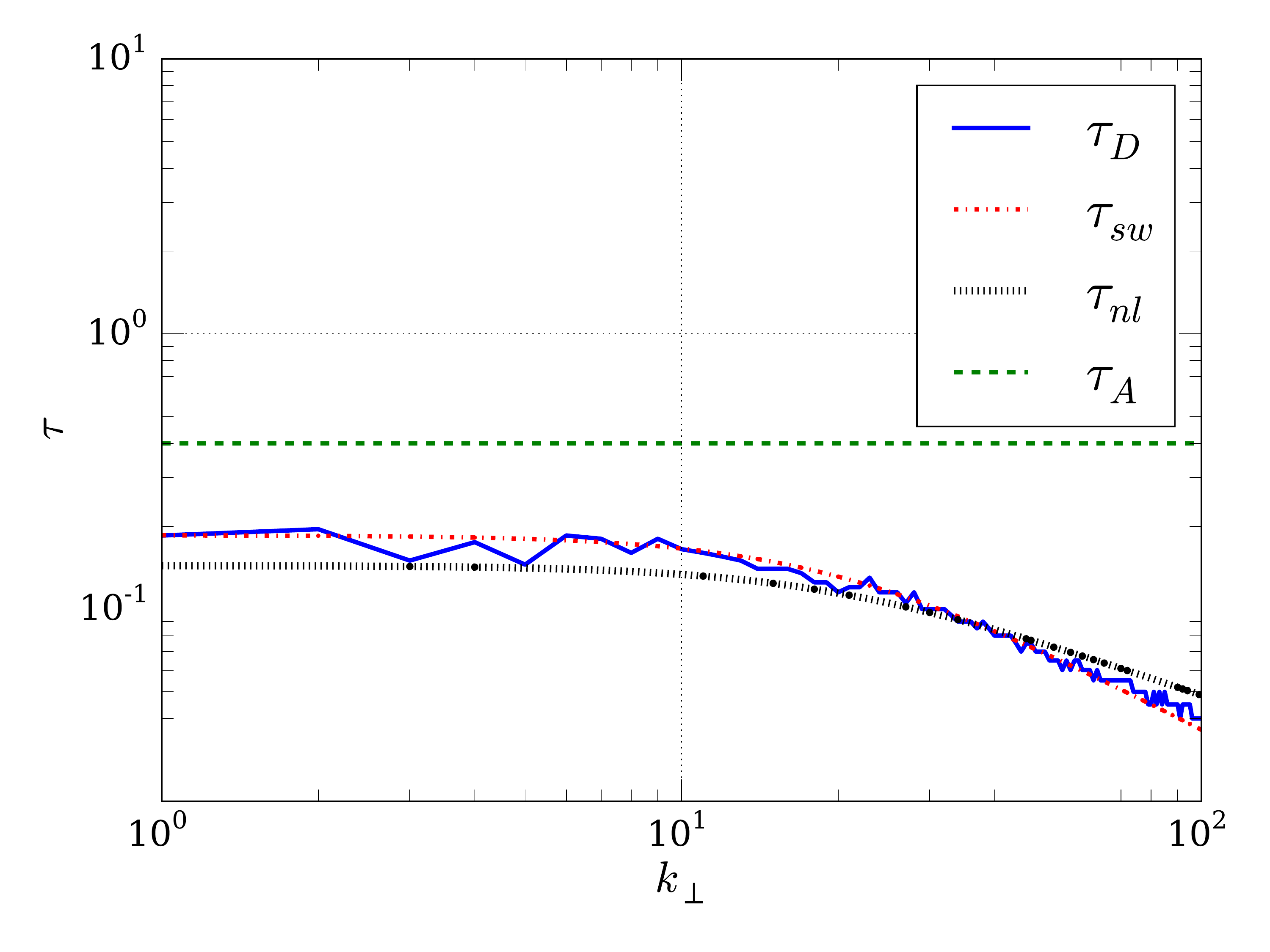}}
  \caption{Decorrelation times $\tau_D$ for the run with
    $B_0=0.25$. In each panel $k_\parallel$ is held constant and
    $k_\perp$ is varied; (a) $k_\parallel = 0$, (b) $k_\parallel =
    10$, and (c) $k_\parallel = 20$. The straight lines indicate 
    theoretical predictions for the scaling of the relevant physical time
    scales. The measured value of $\tau_D$ is always close to
    $\tau_{sw}$.}
  \label{fig5:B025_bvf_b_kpara}
\end{figure}

As mentioned before, in the general case it can be difficult to
differentiate between the effects of sweeping and Alfv\'en
propagation, as both timescales vary as $k^{-1}$. However, in the
anisotropic case (i.e., in the presence of the guide field) we can use
the scaling observed with respect to parallel and perpendicular
wavenumbers to make the distinction possible. In 
Fig.~\ref{fig5:B025_bvf_b_kperp} we employ
results from the $B_0=0.25$ run to compute decorrelation times for
Fourier modes as a function of $k_\parallel$, for several fixed values
of $k_\perp$. Already for this relatively small value of $B_0$ it can
be seen that the observed correlation times are closer to 
the theoretically expected sweeping time than to all the other times
(local nonlinear time or Alfv\'enic time). This is consistent with the
results of the wavenumber and frequency energy spectrum shown
previously in Fig.~\ref{fig3:B025_bvf_Etot_kperp0}. A complementary
view of the same run with $B_0=0.25$ is given in 
Fig.~\ref{fig5:B025_bvf_b_kpara}, which shows the decorrelation time
$\tau$ as a function of $k_\perp$ for several fixed values of
$k_\parallel$. The conclusion is once again that the sweeping time is
controlling $\tau_D$ at all but the largest scales, as only for
$k_\perp=0$ and for $k_\parallel$ between $\approx 1$ and $\approx 4$
$\tau_D$ is closer to the Alfv\'en time.

\begin{figure}
  \centering
  \subfigure[$k_\perp=0$]{\includegraphics[width=1\columnwidth]{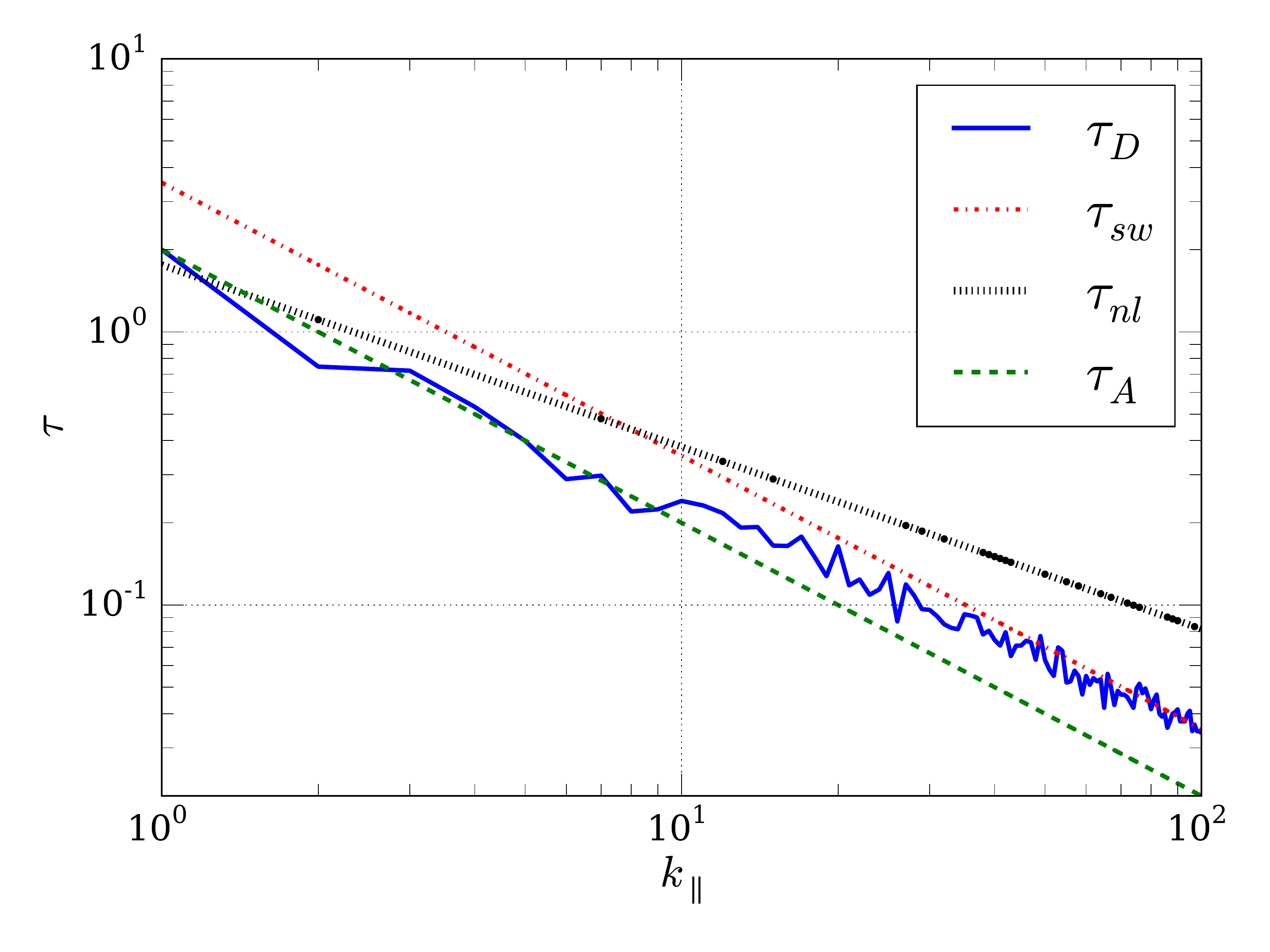}}

  \subfigure[$k_\perp=10$]{\includegraphics[width=1\columnwidth]{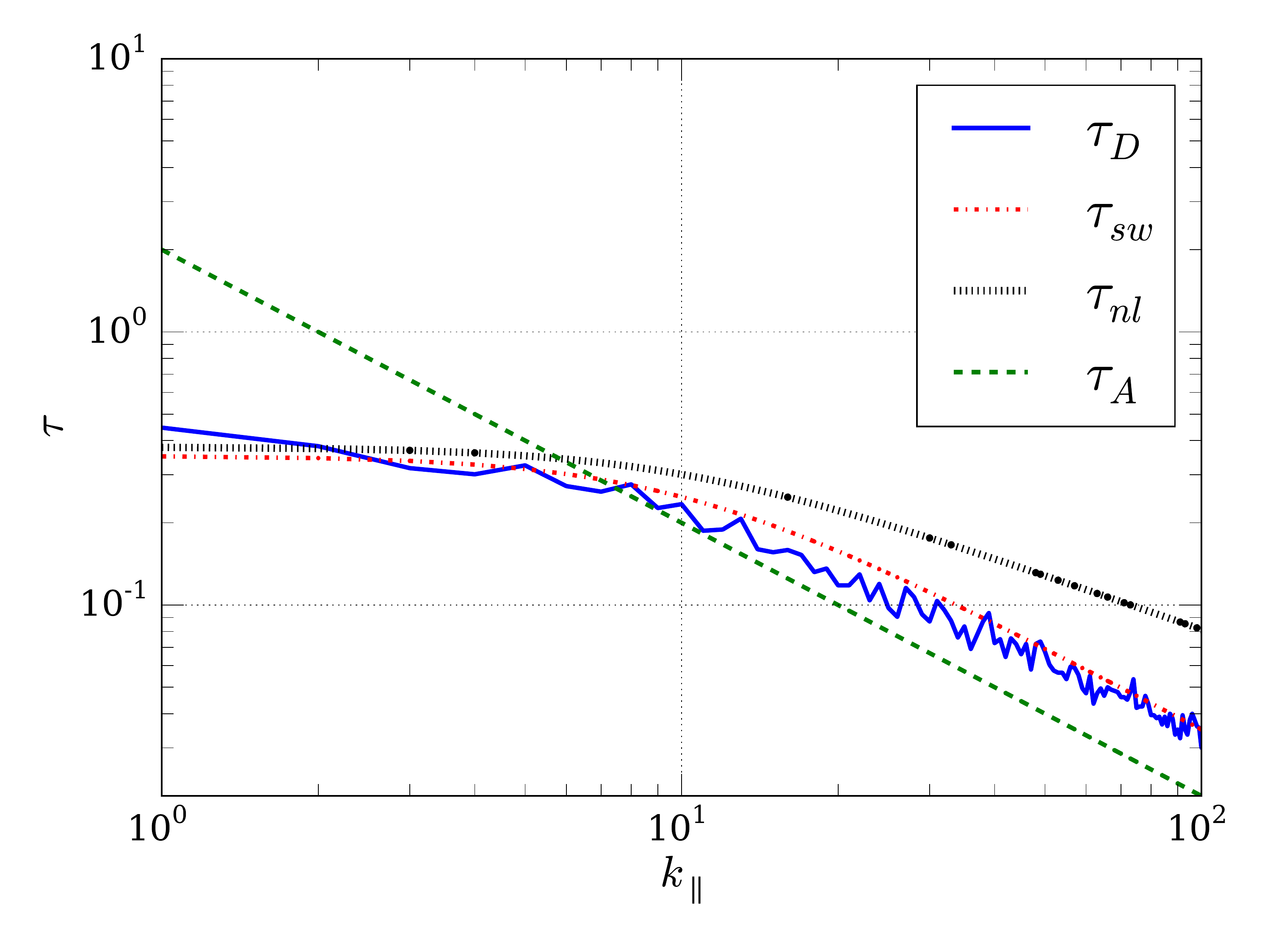}}

  \subfigure[$k_\perp=20$]{\includegraphics[width=1\columnwidth]{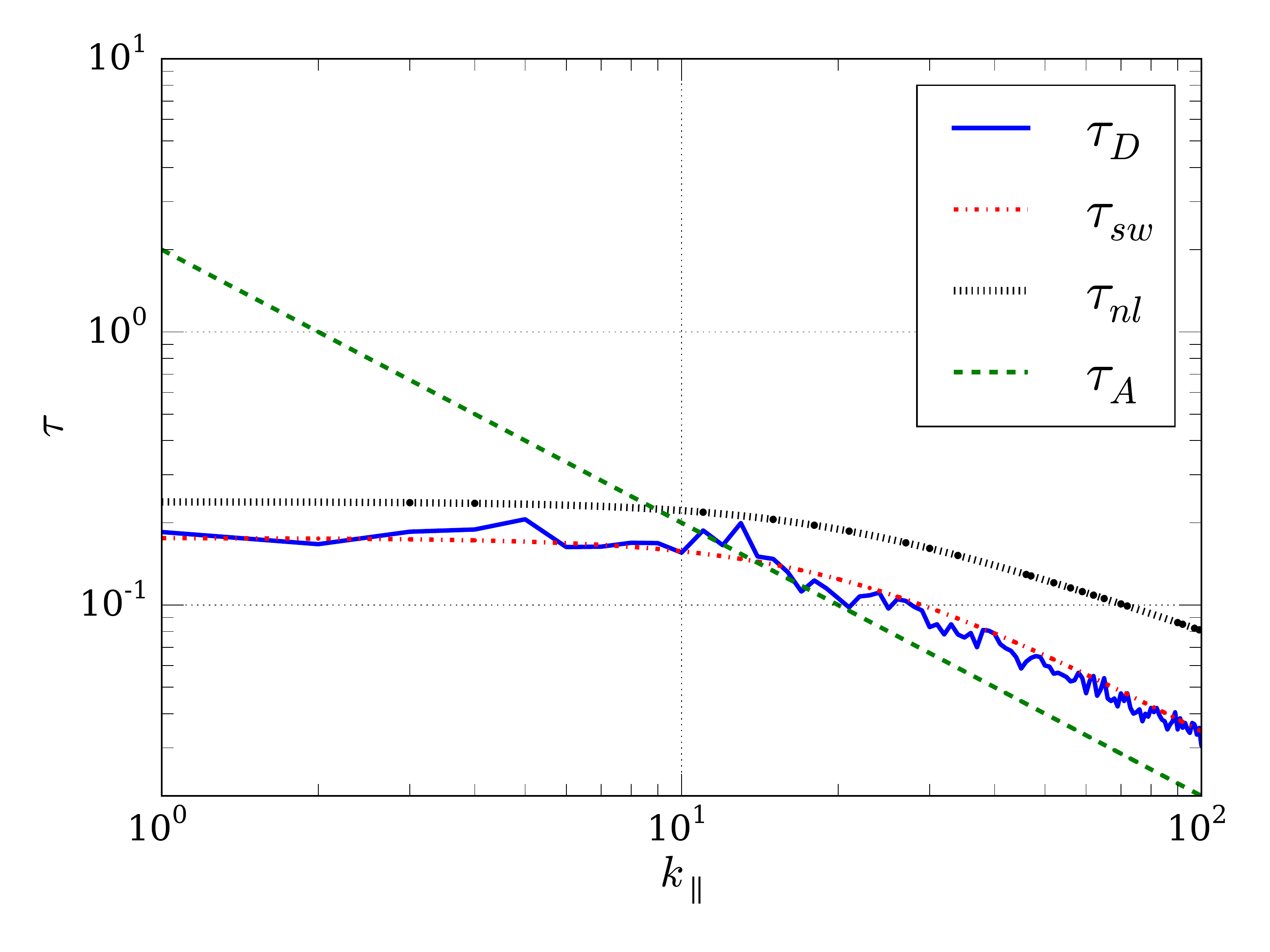}}
  \caption{Decorrelation times $\tau_D$ for the run with $B_0=1$. In each
    panel $k_\perp$ is held constant and $k_\parallel$ is varied; (a)
    $k_\perp=0$, (b) $k_\perp = 10$, and (c) $k_\perp = 20$. The
    straight lines indicate theoretical predictions for
    the scaling of the relevant physical time scales.}
  \label{fig5:B1_bvf_b_kperp}
\end{figure}

\begin{figure}
  \centering
  \subfigure[$k_\parallel=0$]{\includegraphics[width=1\columnwidth]{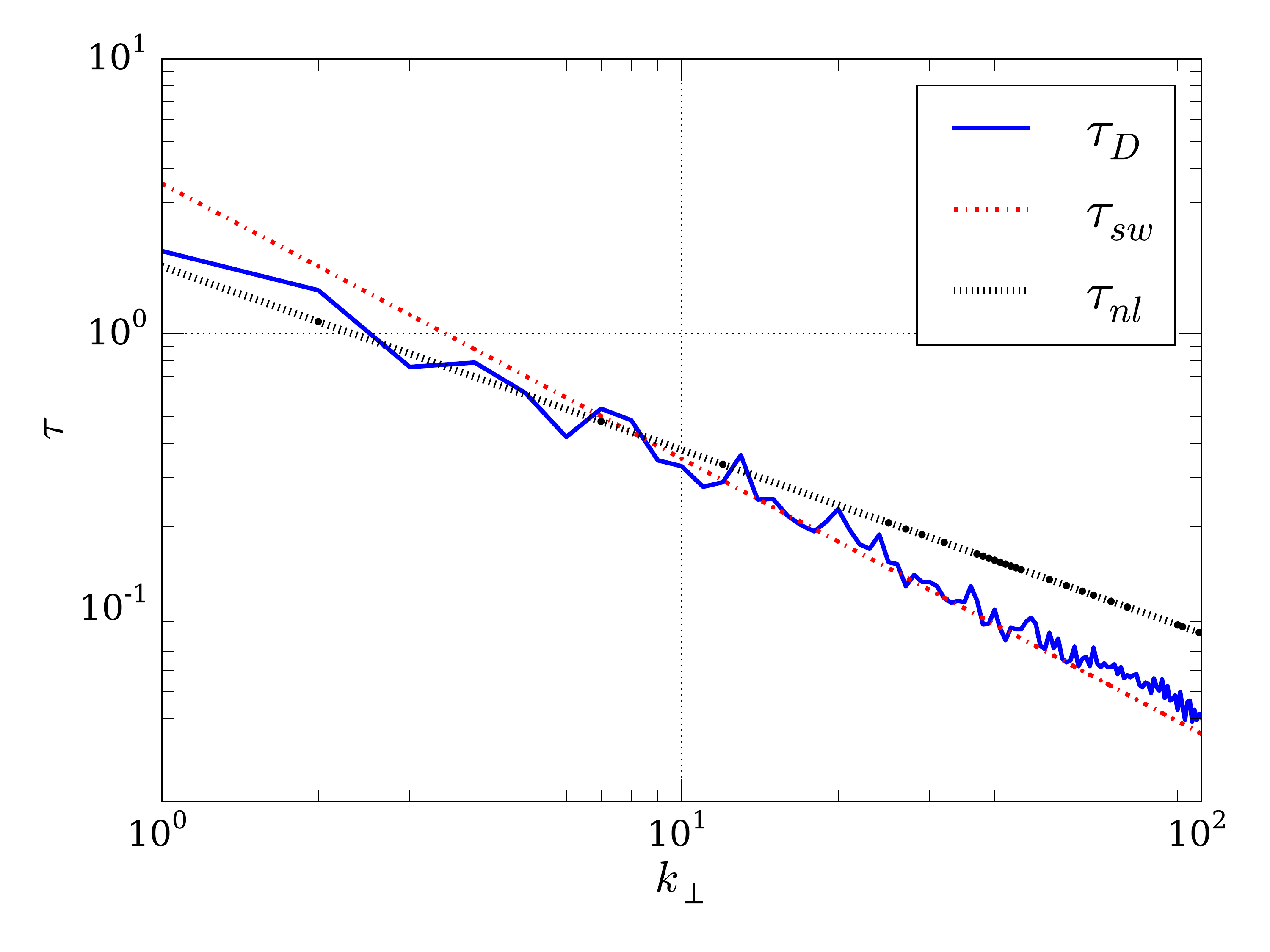}}

  \subfigure[$k_\parallel=10$]{\includegraphics[width=1\columnwidth]{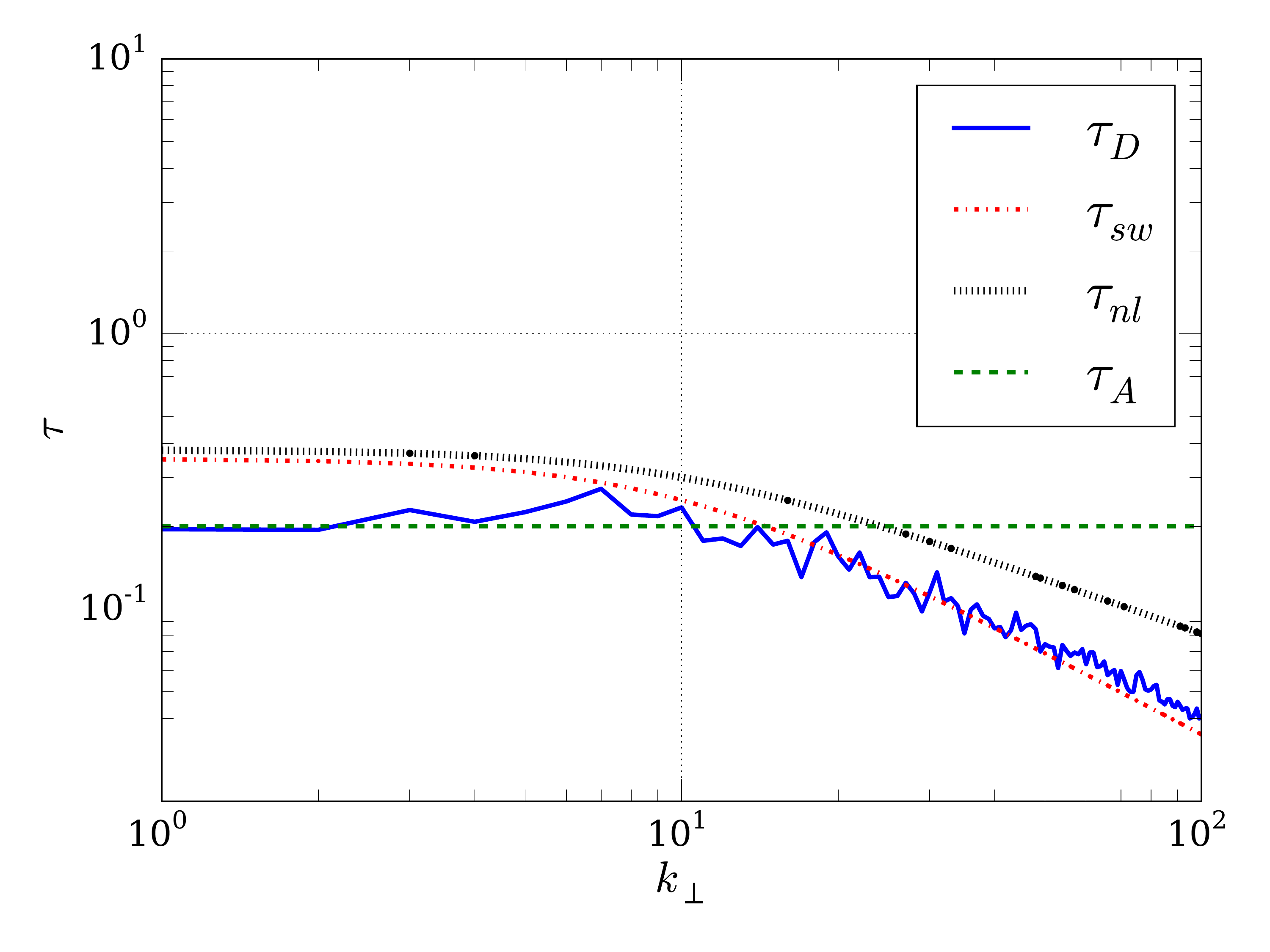}}

  \subfigure[$k_\parallel=20$]{\includegraphics[width=1\columnwidth]{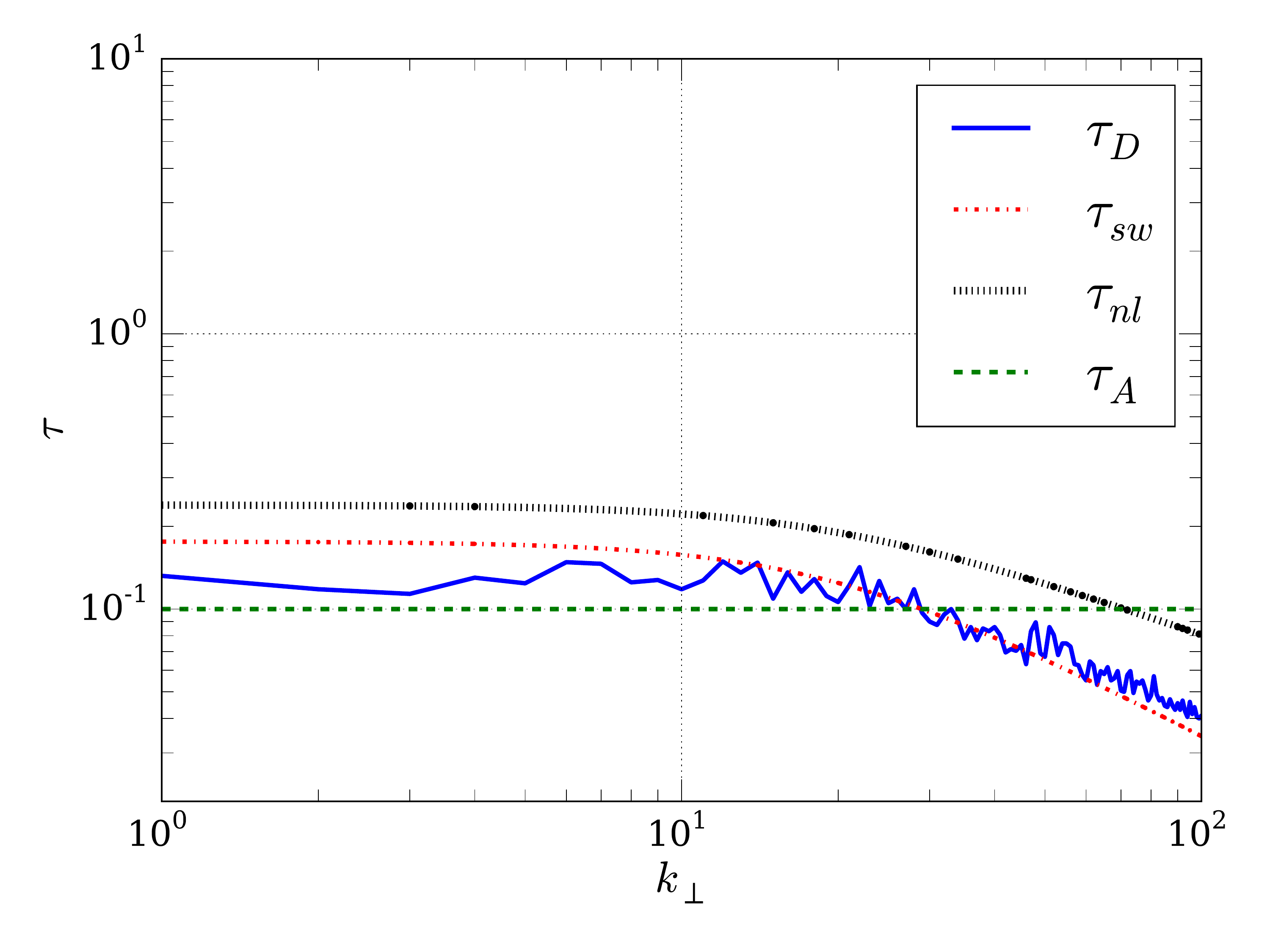}}
  \caption{Decorrelation times $\tau_D$ for the run with $B_0=1$. In each
    panel $k_\parallel$ is held constant and $k_\perp$ is varied; (a)
    $k_\parallel = 0$, (b) $k_\parallel = 10$, and (c) $k_\parallel =
    20$. The straight lines indicate theoretical predictions for
    the scaling of the relevant physical time scales.}
  \label{fig:11} 
\end{figure}

The tendency for time decorrelation to be controlled by sweeping is
again seen in the run with the somewhat stronger mean field $B_0=1$.
These results for the correlation time are shown in in 
Figs.~\ref{fig5:B1_bvf_b_kperp} and \ref{fig:11}. Again, only at low
values of $k_{\parallel}$ and for $k_{\perp}=0$ it can be seen that the
decorrelation time is closer to the Alfv\'enic time. This tendency was
also observed in the wavenumber and frequency spectrum of 
Fig.~\ref{fig3:B1_bvf_Etot_kperp0}.

\begin{figure}
  \centering
  \subfigure[$k_\perp=0$]{\includegraphics[width=1\columnwidth]{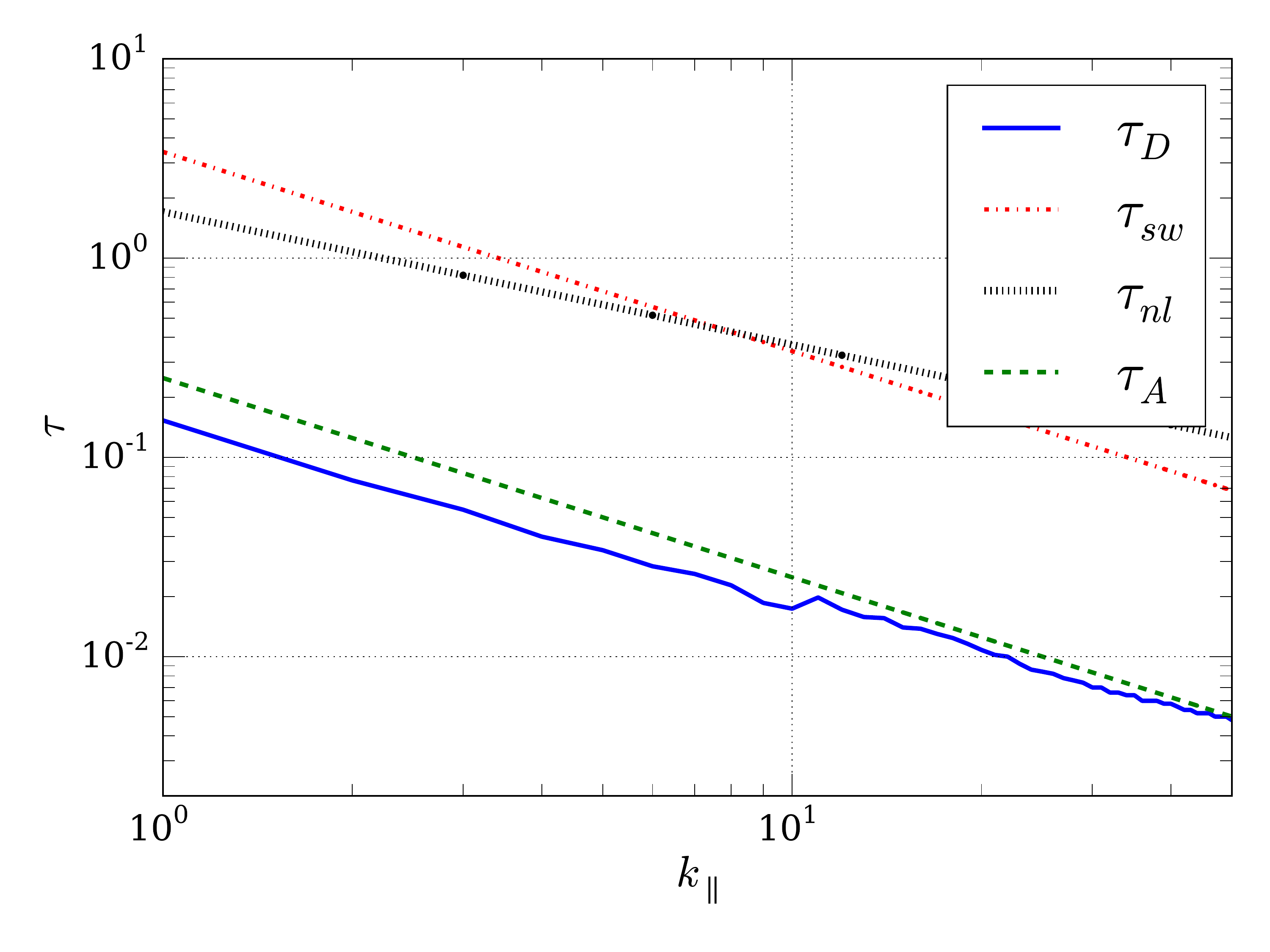}}

  \subfigure[$k_\perp=10$]{\includegraphics[width=1\columnwidth]{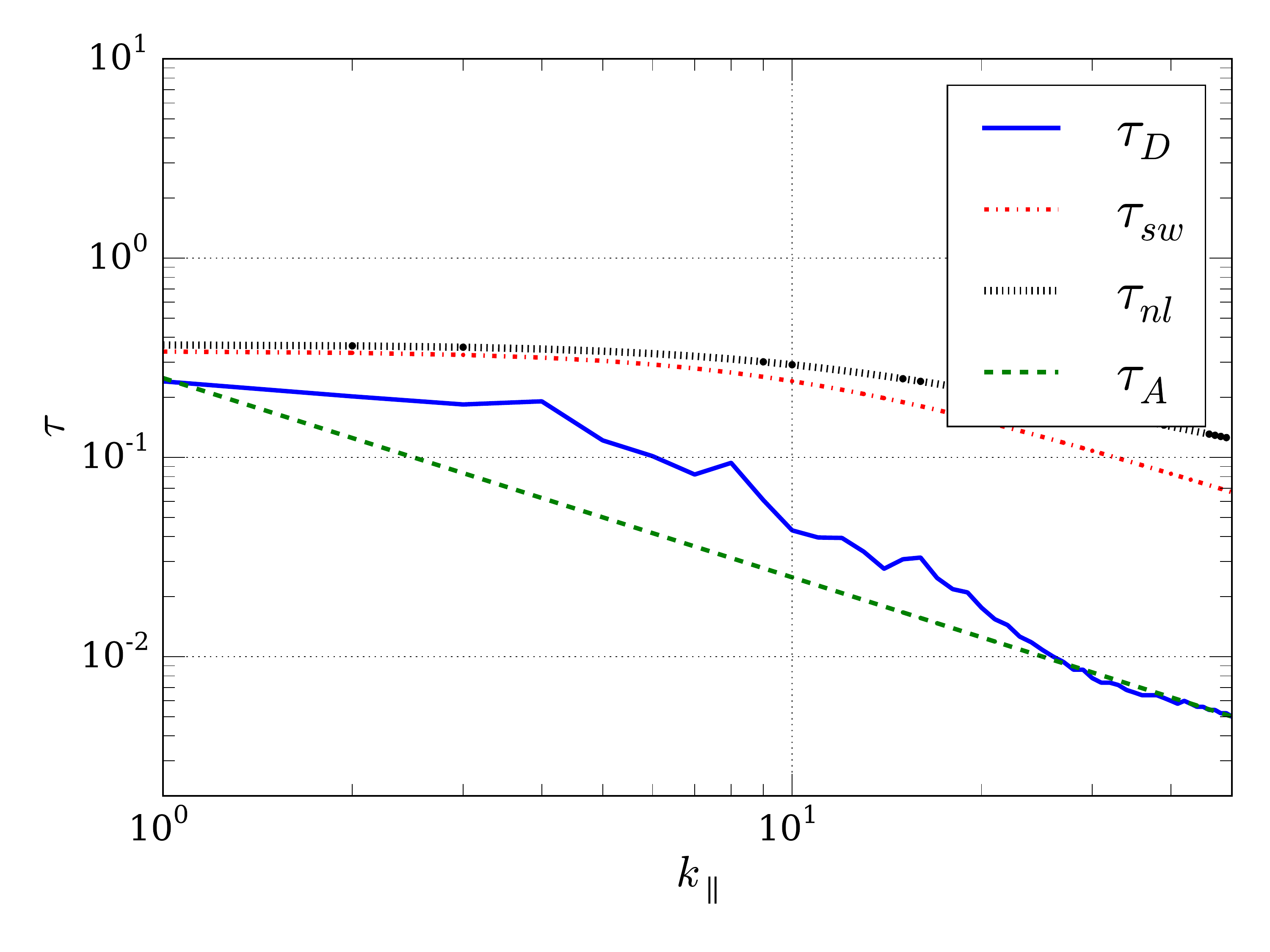}}

  \subfigure[$k_\perp=20$]{\includegraphics[width=1\columnwidth]{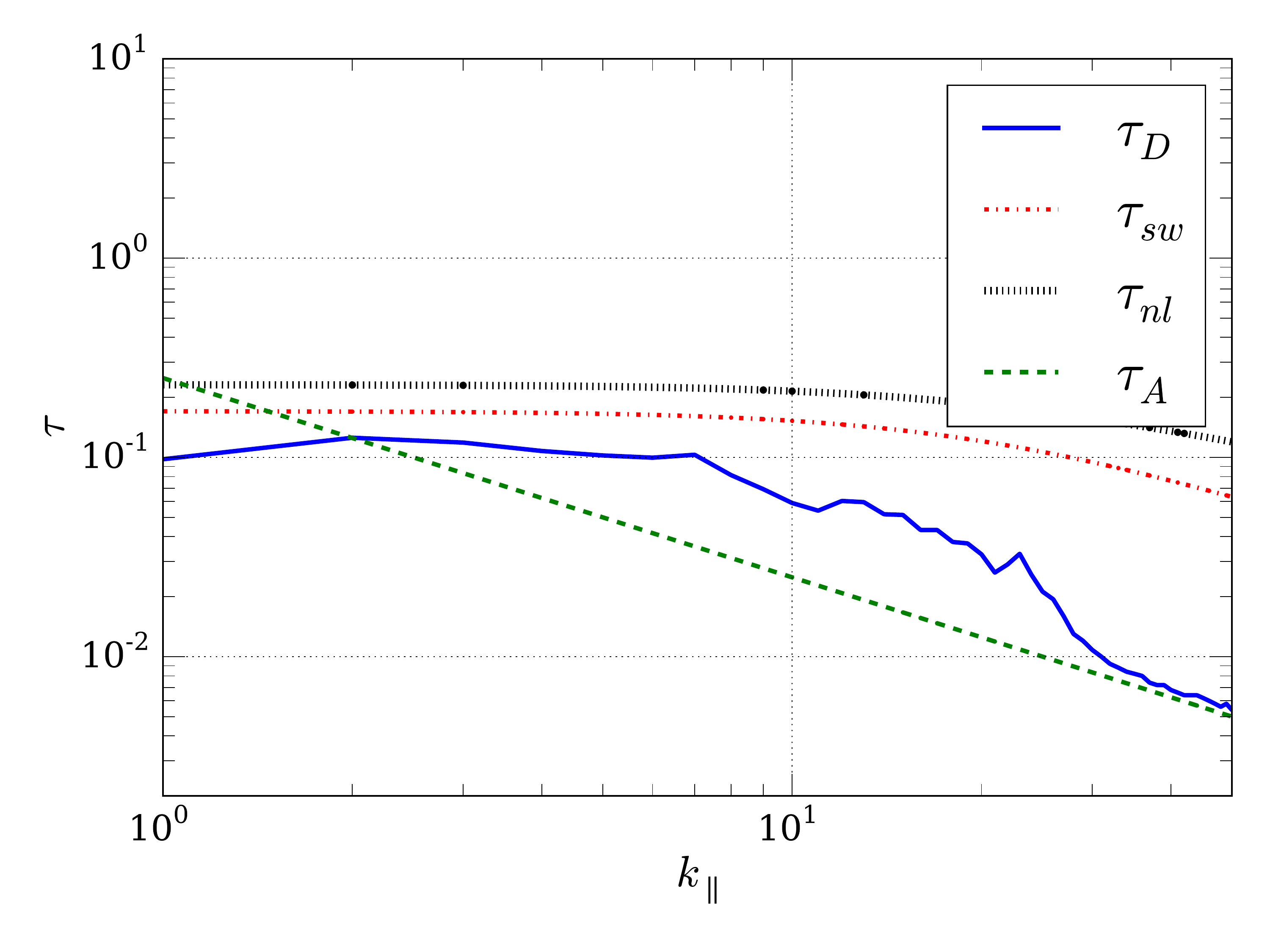}}
  \caption{Decorrelation times $\tau_D$ for the run with $B_0=8$. In each
    panel $k_\perp$ is held constant and $k_\parallel$ is varied; (a)
    $k_\perp=0$, (b) $k_\perp = 10$, and (c) $k_\perp = 20$. The
    straight lines indicate theoretical predictions for
    the scaling of the relevant physical time scales. In this case the
    Alfv\'en time controls the decorrelation at multiple wavenumbers.}
  \label{fig5:B8_bvf_b_kperp}
\end{figure}

\begin{figure}
  \centering
  \subfigure[$k_\parallel=0$]{\includegraphics[width=1\columnwidth]{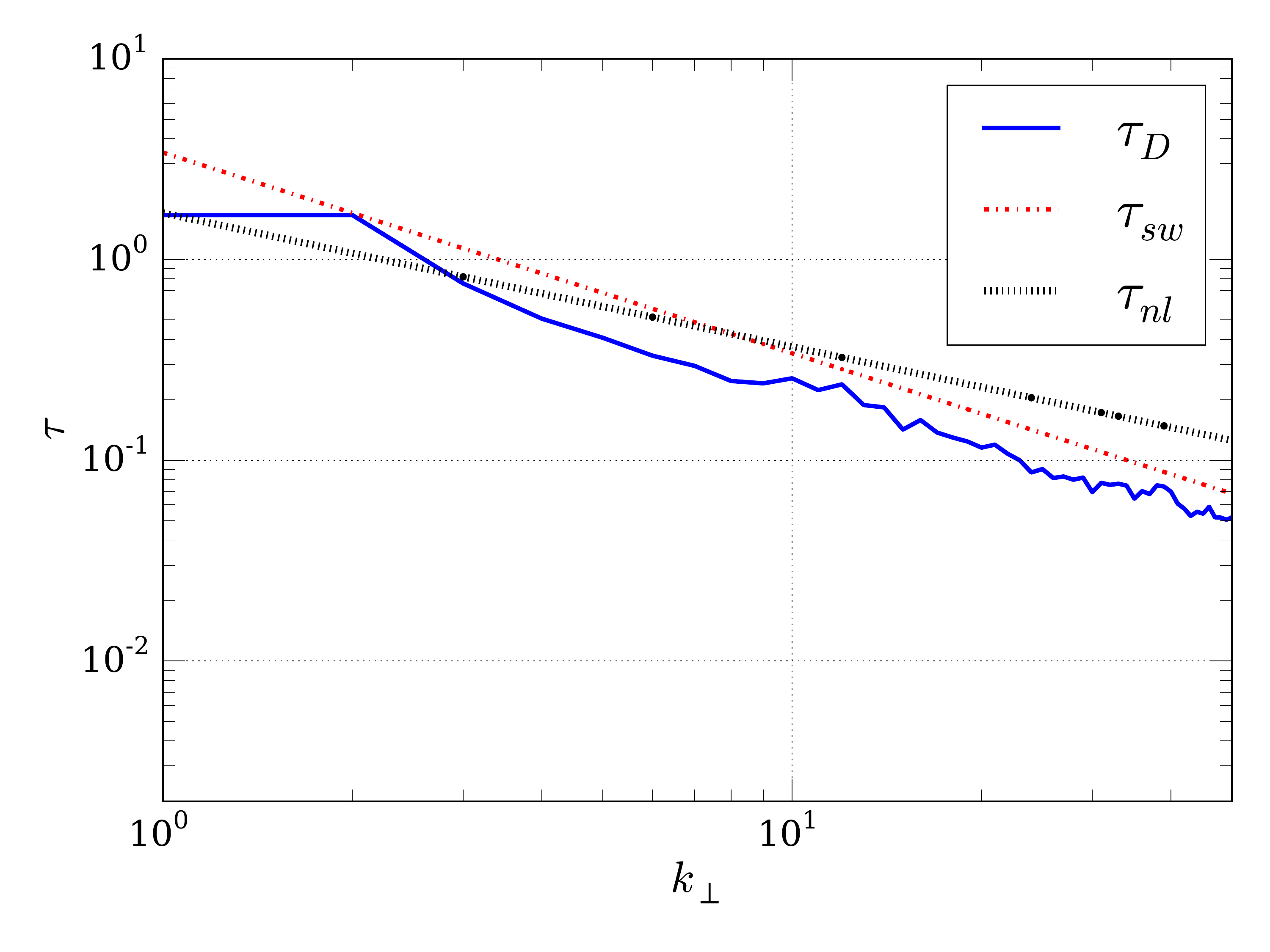}}

  \subfigure[$k_\parallel=10$]{\includegraphics[width=1\columnwidth]{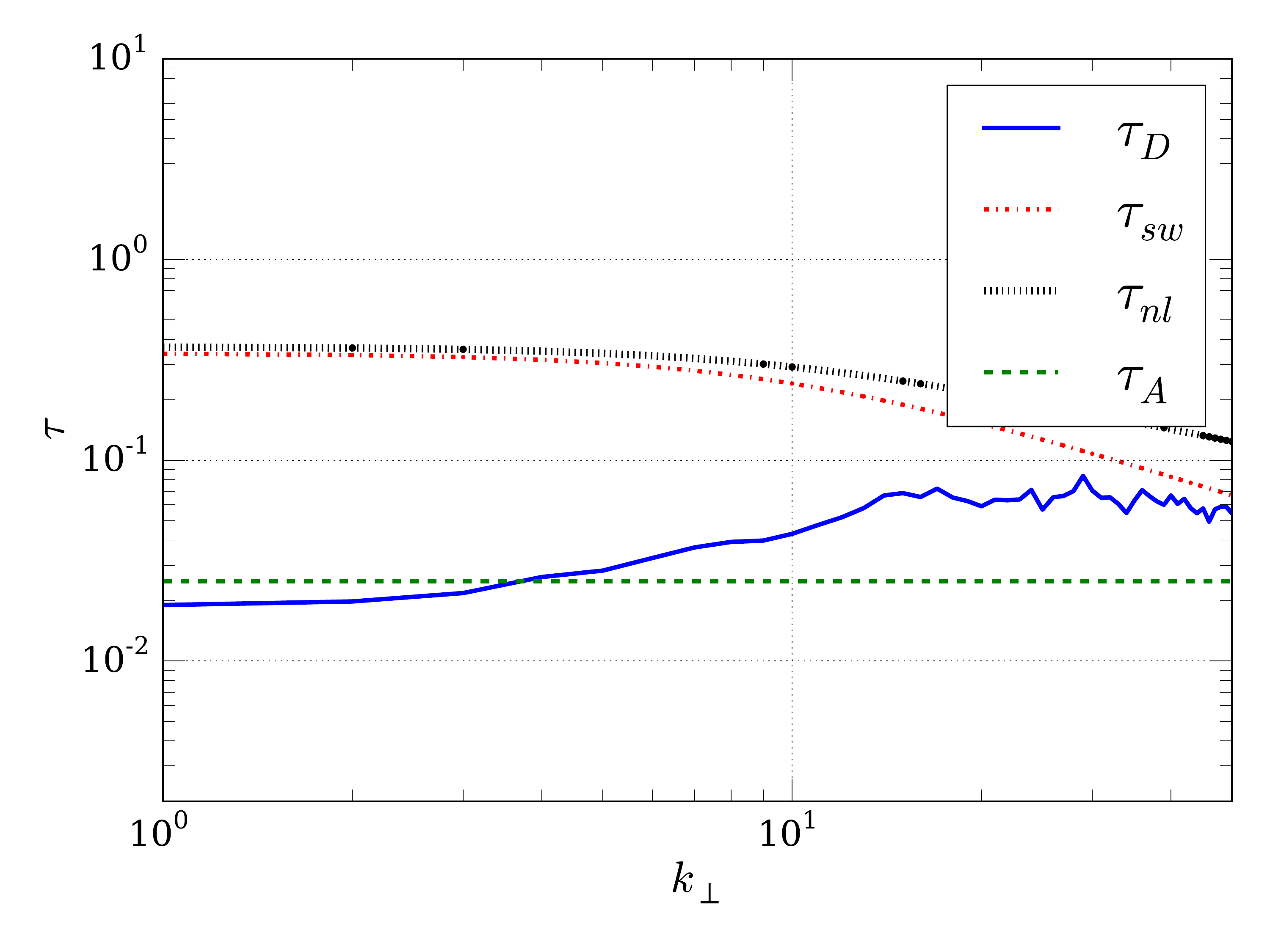}}

  \subfigure[$k_\parallel=20$]{\includegraphics[width=1\columnwidth]{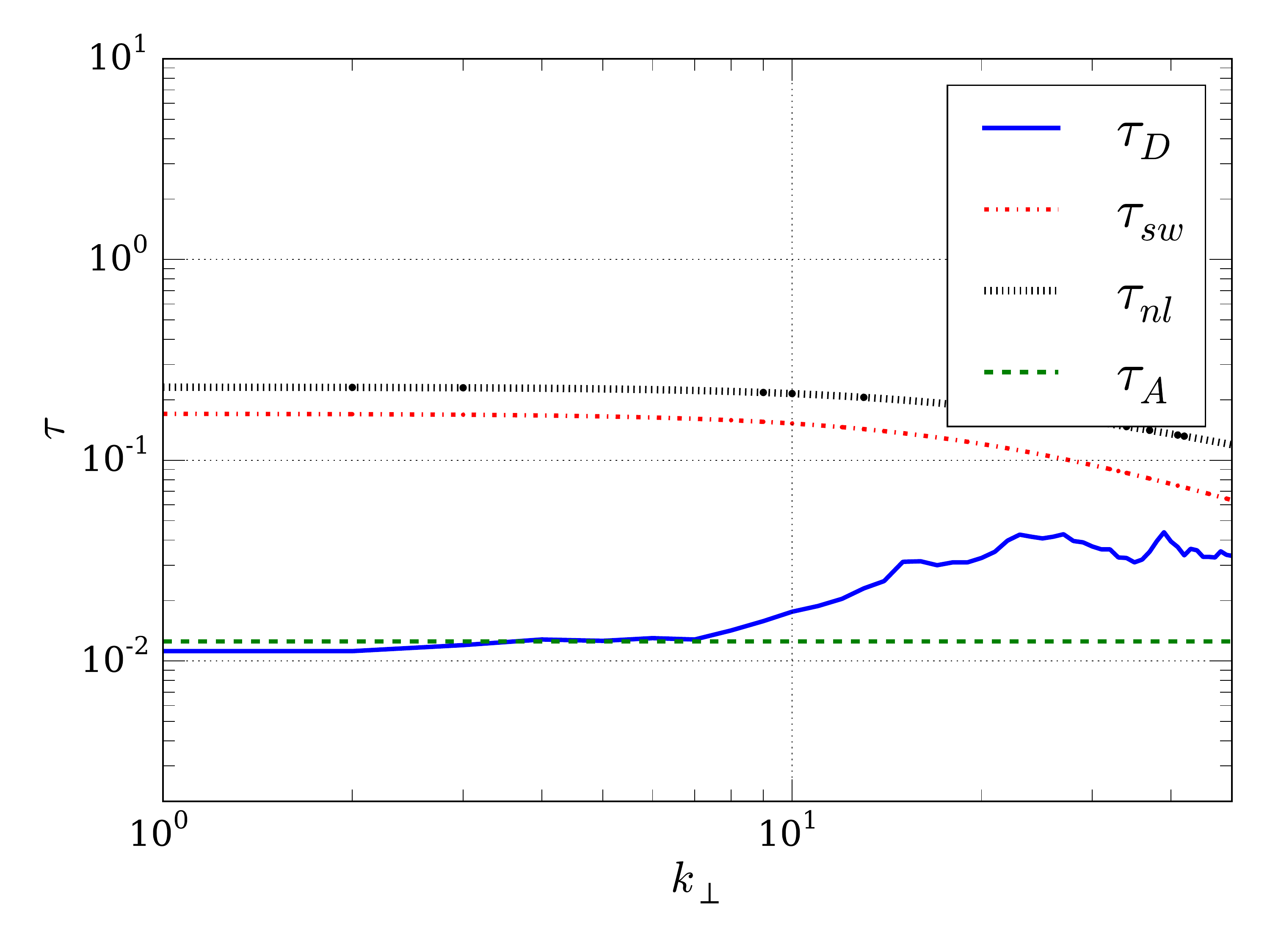}}
  \caption{Decorrelation times $\tau_D$ for the run with $B_0=8$. In each
    panel $k_\parallel$ is held constant and $k_\perp$ is varied; (a)
    $k_\parallel = 0$, (b) $k_\parallel = 10$, and (c) $k_\parallel =
    20$. The straight lines indicate theoretical predictions for
    the scaling of the relevant physical time scales. The Alfv\'en
    time controls the decorrelation up to $k_\parallel \approx 10$.}
  \label{fig5:B8_bvf_b_kpara}
\end{figure}

Finally, we analyze the behavior of the decorrelation time $\tau$ for
the run with the largest mean magnetic field value that we
considered, $B_0=8$. The results are presented in 
Figs.~\ref{fig5:B8_bvf_b_kperp} and \ref{fig5:B8_bvf_b_kpara}, analyzed in
the same way as in the previous two cases.  For low values of $k_\perp$
one finds that the Alfv\'enic time dominates the decorrelations
(approximately up to $k_\parallel = 10$, see
Fig.~\ref{fig5:B8_bvf_b_kpara}). For larger values of $k_{\perp}$,
however, the decorrelation time departs from the Alfv\'en time and
slowly approaches the sweeping time scale. This is consistent with the
spatio-temporal spectrum in Fig.~\ref{fig3:B8_bvf_Etot_kperp0}, which
concentrated energy near the Alfv\'en dispersion relation for small
wavenumbers, but broadened towards the sweeping frequencies for large
wave numbers. As a result, it is the competition between these two
time scales that for large values of $B_0$ seems to be responsible for
the broadening of the spatio-temporal spectrum. As long as the
Alfv\'en time is much faster than other time scales in the system, the
flow excites Alfv\'en waves which dominate the mode decorrelation. But
as other time scales approach the time scale of the waves (or become
faster, as it happens for smaller values of $B_0$), the system
switches the dominant time scale in the decorrelation.

\section{Conclusions}\label{sec_Conclusions}

In this paper we have studied the time correlations that enter into
magnetohydrodynamics in the incompressible approximation.  Even in the
simpler case of hydrodynamics one expects both space and time
correlations to be relevant to the physics of turbulence, as these
independent properties can be embodied in the two point, two time
correlation $R_{ij}({\bf r},t)$ tensor, e.g., a straightforward
generalization of Eq.~(\ref{eq:Rbij}).  Analogous correlations may
also be written for the components of vector fluid velocity ${\bf u}$
and other quantities.  The spatial transform of the correlation (or,
equivalently the second order spatial structure functions) at zero
time lag $\tau$ provides information about the spatial distribution of
energy over scales.  Accordingly the zero spatial lag correlation,
evaluated at varying time and transformed to frequency, provides
analogous information about the distribution of energy over time
scales.  Here we studied the correlations in time for a given
wavenumber or spatial scale for the magnetohydrodynamics model.

The MHD case is more complex than hydrodynamics because two basic
fields are involved -- velocity and magnetic field. Also because a
mean magnetic field is not removed by a Galilean transform, while a
mean velocity can be removed in this way. The mean magnetic field
therefore imposes a preferred direction.  In addition, MHD possesses a
new and anisotropic wave mode, the Alfv\'en mode, that introduces the
possibility of spectral and correlation anisotropy, as well as a new
times scale, the Alfv\'en time.  Because of these effects the analysis
of time decorrelation also become more complex, with at least three
time scales to examine -- Alfv\'en, sweeping and nonlinear scales --
as well as potential for anisotropy of the decorrelation rate.

Both random sweeping and Alfv\'enic correlation are non-local effects,
in the sense that they couple the large scales with relatively smaller
length scales. The results shown here support the conclusion that
non-local effects (in spectral space) play an important role in MHD
turbulence (in agreement with studies considering shell-to-shell
transfers \cite{alexakis_turbulent_2007, alexakis_anisotropic_2007,
  teaca_energy_2009, mininni_scale_2011}), and that decorrelations are
mainly dominated by the sweeping and Alfv\'enic interactions,
confirming previous studies of isotropic MHD
\cite{servidio_time_2011}. However, compared with the previous
studies, the analysis presented here can further distinguish between
sweeping and Alfv\'enic effects, and the results support the
conclusion that the sweeping interaction dominates the decorrelation
for moderate values of $B_0$, while for large values of the mean field
$B_0$ and at large scales (low perpendicular wavenumbers) the
decorrelations are more controlled by the Alfv\'enic interactions.
The relevant interactions are the Alfv\'en waves, and as such it can
be concluded that waves are still present in MHD turbulence and
dominate the decorrelations essentially for parallel wavenumbers
(aligned with the mean field, see also \cite{meyrand_direct_2016,
  meyrand_weak_2015}). Our results further indicate that the system
selects, in effect, the shortest decorrelation time available.  A
simple and relevant construct is that the rate of decorrelation is the
sum of the rates associated with each relevant time scale (see, e.g.,
\cite{zhou_magnetohydrodynamic_2004}). As a result, even for large
values of the guide field $B_0$, for sufficiently small scales in
which the sweeping time becomes faster than the Alfv\'enic time, after
a broad range of scales dominated by Alfv\'en waves the system
transitions to a sweeping dominated behaviour.

It is of interest to recall that the relevant time decorrelation
associated with energy transfer in turbulence is not the Eulerian time
correlation that we have considered (fixed spatial point, varying
time), but rather the Lagrangian time decorrelation, computed
following a material fluid element.  In this regard, it is well known
that neither sweeping nor Alfv\'enic wave propagation can directly
produce spectral transfer in idealized homogeneous models.  In part
due to these complications, no complete theory exists at present that
links the spatial correlation and the time correlations in MHD or
hydrodynamic turbulence.  On the other hand it is clear that in MHD,
both sweeping and Alfv\'en wave propagation contribute to the total
time variation at a point (Eulerian frequency spectrum), and are
therefore influential in limiting prediction.  These time scales are
also important features in understanding the scattering of charged
test particles, such as low energy cosmic rays
\cite{bieber_proton_1994}, as well as in accounting for the
distribution of accelerations, which is related to intermittency
\cite{nelkin_time_1990}.

The observed behavior of MHD time decorrelation, exemplified by the
new results presented here, thus have applications in a number of
subjects, including charged particle scattering theory
\cite{schlickeiser_cosmic-ray_1993, nelkin_time_1990}, interplanetary
magnetic field and magnetospheric dynamic \cite{miller_critical_1997},
and interpretation of spacecraft data from historical and future
missions \cite{matthaeus_ensemble_2016}.  Looking towards future
prospects, we note that there has been some success in establishing
empirical connections between the sweeping time scale to the observed 
Eulerian time decorrelation in hydrodynamics
\cite{chen_sweeping_1989}.  Similar ideas for MHD (e.g.,
\cite{matthaeus_dynamical_1999}) might be exploited to better
understand, or at least empirically model, the relationship in MHD
between spatial structure and time decorrelation, an effort that would
directly benefit from the novel results presented here.

\section{Acknowledgments}\label{sec_Acknowledgments}

R.L., P.D., and P.D.M.~acknowledge support from the grants UBACyT
No.~20020110200359 and 20020100100315, and PICT No.~2011-1529,
2011-1626, and 2011-0454.

W.H.M. is partially supported by NASA LWS-TRT grant NNX15AB88G, NASA
Grant NNX14AI63G (Heliophysics Grand Challenge Theory), and the Solar
Probe Plus mission through the Southwest Research Institute ISOIS
project D99031L.

\bibliography{Bibliografia}
\bibliographystyle{unsrt}

\end{document}